\documentclass[prc,onecolumn,showpacs,nofootinbib]{revtex4}
\usepackage{amsmath,amssymb,color,bm,graphicx}
\newcommand{\al}{\alpha}

\newcommand{\De}{\Delta}

\newcommand{\eps}{\epsilon}

\newcommand{\om}{\omega}

\newcommand{\ka}{\kappa}

\newcommand{\dsp}{\displaystyle}
\newcommand\eqn[1]{(\ref{#1})}      % parentheses around the LaTex "ref" macro
\newcommand{\beq}{\begin{equation}}
\newcommand{\eeq}{\end{equation}}
\newcommand{\be}{\begin{equation}}
\newcommand{\ee}{\end{equation}}
\newcommand{\ba}{\begin{array}}
\newcommand{\ea}{\end{array}}
\newcommand{\bc}{\begin{center}}
\newcommand{\ec}{\end{center}}
\newcommand{\bea}{\begin{eqnarray}}
\newcommand{\eea}{\end{eqnarray}}
\newcommand{\bi}{\begin{itemize}}  %\setlength{\itemsep}{0\parsep}}
\newcommand{\ei}{\end{itemize}}
\newcommand{\ben}{\begin{enumerate}} %\setlength{\itemsep}{0\parsep}}
\newcommand{\een}{\end{enumerate}}

\newcommand\hide[1]{}

\renewcommand{\Re}{{\rm Re}\,}

\usepackage[normalem]{ulem}  % \sout{old text} for strikeout

\newcommand{\Qt}{{\tilde Q}}\newcommand{\X}{X}

\definecolor{red}{rgb}{0.8,0,0}
\definecolor{violet}{rgb}{0.4,0,0.4}
\definecolor{green}{rgb}{0,0.5,0.0}
\definecolor{navy}{rgb}{0.0,0.0,0.6}
\definecolor{orange}{rgb}{0.8,0.2,0.0}
\definecolor{blue}{rgb}{0.3,0.0,0.8}

\begin{document}

\title{Transport coefficients of two-flavor superconducting quark matter}
\author{Mark G. Alford}
\email{alford@wuphys.wustl.edu}
\affiliation{Physics Department, Washington University,
St.~Louis, MO~63130-4899, USA}
\author{Hiromichi Nishimura}
\email{nishimura@physik.uni-bielefeld.de}
\affiliation{Faculty of Physics, University of Bielefeld, D-33615 Bielefeld, Germany}
\author{Armen Sedrakian}
\email{sedrakian@th.physik.uni-frankfurt.de}
\affiliation{Institute for Theoretical Physics, 
J. W. Goethe-University, D-60438 Frankfurt am Main, Germany}

\begin{abstract}
\begin{description}
\item[Background:]
The two-flavor color superconducting (2SC) phase of quark matter is a
possible constituent of the core of neutron stars. To assess its impact
on the observable behavior of the star one must analyze
transport properties, which
in 2SC matter are controlled by the scattering of gapless fermionic
modes by
each other and possibly also by color-magnetic flux tubes.
\item[Purpose:] We determine the electrical and thermal conductivities
and the shear viscosity of 2SC matter.
\item[Methods:] We use a variational formulation of transport theory,
treating the strong and electromagnetic interactions via a weak coupling
expansion.
\item[Results:]  We provide the leading order
scaling of the transport coefficients with temperature and chemical
potential
as well as accurate fits to our numerical results.  We also find that
the
scattering of fermions by color-magnetic flux tubes
is insignificant for thermal conductivity, but may 
contribute to the electrical conductivity
and shear viscosity in the limit of very low temperature or
high magnetic field. We also estimate the transport coefficients in
unpaired quark matter.
\item[Conclusions:] Our calculation has set the stage for exploration of
possible signatures of the presence of 2SC quark matter in neutron
stars.
\end{description}
\end{abstract}

\date{\today} 
% hardwire the date so arXiv doesn't change it

%----------------------- PACS ----------------------------%
%97.60.Jd  Neutrons stars                                 %
%21.65.Qr  Quark matter                                   % 
%12.38.Mh  Quark-gluon plasma in quantum chromodynamics   %
%47.32.C-  Vortex dynamics (fluid flow)                   % 
%74.25.Uv  Vortex phases (superconductivity)              % 
%03.65.Ta  Aharonov-Bohm effect quantum mechanics         % 
%67.10.Jn  Transport processes in quantum fluids          %
%---------------------------------------------------------%


\maketitle
\section{Introduction}
\label{sec:intro}

Transport coefficients of dense matter play a central role in the
modeling of astrophysical phenomena in compact stars.  The thermal and
magnetic evolution of compact stars, their rotational dynamics, and
emission of electromagnetic and gravitational waves, all depend on the
transport properties of different phases of dense matter.  
 
In the core of a massive compact star, gravity compresses matter to a density where it may undergo a transition to quark matter which at
sufficiently low temperature should be in one of the color superconducting
phases~\cite{Bailin:1983bm,Alford:2007xm}. Transport in a given
phase is determined by the low-energy excitations of that phase, which are
controlled by the symmetry breaking pattern.
At asymptotically high density the favored phase is
the color-flavor-locked (CFL) phase, where all the quark
flavors and colors form Cooper pairs with zero total
momentum~\cite{1999NuPhB.537..443A}.  
The only excitations of the CFL phase at low
temperature are superfluid phonons, whose interactions
determine the transport coefficients of this
phase~\cite{Shovkovy:2002kv,2007JCAP...08..001M,2005JHEP...09..076M}. The nature of quark
pairing at lower densities, which may include the range relevant
for compact stars, remains uncertain.
One candidate is the two-flavor color-superconducting (2SC) phase,
in which up ($u$) quarks and down ($d$)
quarks pair in a color antitriplet state leaving one of the colors 
% (say blue)
unpaired \cite{Bailin:1983bm,Alford:2007xm}.  
In this paper we calculate key transport properties of this phase.

The paper is structured as follows.  In Sec.~\ref{sec:ff_scattering} we discuss the relevant interactions among the ungapped fermions and calculate the scattering matrix
elements for the fermions interacting via exchange of gauge bosons in the 2SC phase.
Section \ref{Multicomponent}
develops a general formalism for transport in multicomponent
systems starting from the Boltzmann equation. 
After briefly explaining the physics of transport in the 2SC phase in Sec.~\ref{sec:qualitative} and our approximation schemes in Sec.~\ref{sec:approximation}, 
we go on to compute the electrical conductivity, thermal conductivity, and
shear viscosity of 2SC matter (Secs.~\ref{sec:electrical_conductivity}, \ref{sec:thermal_conductivity},
and \ref{sec:shear_viscosity}, respectively).
In Sec.~\ref{sec:fluxtubes} we
compare the fermion-fermion scattering contribution 
to the fermion--flux-tube scattering contribution and
identify the domain where the latter could become important. Our
results are summarized in Sec.~\ref{sec:conclusions}.   We use
``Heaviside-Lorentz'' natural units with $\hbar = c = k_B =
\eps_0 = 1$, where $k_B$ is the Boltzmann constant and $\eps_0$ is the
vacuum permittivity; the electric charge $e$ is related to the fine
structure constant by $\al=e^2/(4\pi)=1/137$ and similarly, the QCD coupling constant $g$ by $\alpha_s = g^2/(4\pi)$.

\section{Fermion-fermion scattering in the 2SC phase}
\label{sec:ff_scattering}

\subsection{Relevant excitations}
\label{sec:relevant_excitations}

The excitations that transport momentum and energy in 2SC
superconductors are ungapped fermions. We fix to unitary gauge, where
the 2SC condensate is uniform over all space and time, and use the
standard convention that the condensate points in the red-green
direction in color space.  The red and green quarks, because of their
2SC pairing, have gaps $\Delta$ which are expected to be around
10\,MeV or larger \cite{Alford:2007xm,Rischke:2000cn} so their occupation is
Boltzmann suppressed and they are frozen out of transport processes at
temperatures appropriate to neutron stars.  This leaves electrons and
blue quarks as the only ungapped fermionic excitations. We will
neglect muons and strange quarks. If they were present, their
Fermi momenta and available phase space would be much smaller, so they
would play a subleading role in transport processes. The main effect
of strange quarks would be to reduce the electron population, 
affecting the dominance of electrons in transport.

% a condition that guarantees that the color superconductor is
% type-II~\cite{Alford:2010qf} and therefore color magnetic flux tubes are formed.
The 2SC phase breaks no global symmetries, so there are no massless
Goldstone bosons, or light pseudo-Goldstone bosons.
The gauge bosons available to mediate fermion-fermion interactions
in the 2SC phase are the eight gluons (generators $T_A$) and the photon 
(generator $Q$). However, because the transport coefficients are determined by the interactions only among the electron ($e$), the blue up quark ($bu$), and the blue down quark ($bd$), most of the gauge bosons can be neglected.
In the 2SC phase, the gauge symmetry breaking pattern is 
$SU(3)_{\rm color} \otimes U(1)_Q \to SU(2)_{rg} \otimes U(1)_\Qt$ \cite{Alford:1997zt,Alford:1999pb}.
The unbroken $SU(2)_{rg}$ symmetry consists of color rotations 
involving the red and green colors. 
The unbroken $U(1)_\Qt$ gauge symmetry is a linear combination of
the original electromagnetic and color symmetries, called ``rotated
electromagnetism,'' generated by $\Qt$ which is a linear combination of $Q$
and the 8th color generator $T_8$. 
The other linear combination of the gauge bosons is called the $X$ boson, which is massive. 
The remaining gauge bosons are irrelevant because they do not couple to electrons, which have no color, and they cannot mediate interactions
between blue quarks, because they all carry some non-blue color.
As we will see, the transport properties of the 2SC phase are
determined by the $\Qt$ interaction, which is weakly-coupled but
long-ranged (Landau damped), and the $T_8$ and $X$ interactions, which are
strongly coupled but short-ranged due to Debye and Meissner screening, respectively.

Lastly, we here compute the chemical potential of each flavor in the 2SC phase. The
symmetry group of massless two-flavor QCD is
$SU(3)_{\rm color}\times SU(2)_L \times SU(2)_R \times U(1)_B$, where 
electromagnetism (generated by a combination of baryon number
and isospin) and color are gauged.
The relevant chemical potentials are $\mu_q$ (coupled to quark number),
$\mu_e$ (coupled to negative electric charge), and
$\mu_3$ and $\mu_8$
which are coupled to the Cartan generators of $SU(3)_{\rm color}$,
$T_3= \mbox{diag} (1/2,-1/2,0) $ and $T_8 =\mbox{diag} (1/(2 \sqrt{3}),1/(2 \sqrt{3}),-1/ \sqrt{3})$. 
Because the color subgroup $SU(2)_{rg}$, which rotates red and green
quarks, is unbroken in the 2SC phase, we have $\mu_3=0$.
The chemical potentials for blue up and blue down quarks are then
\begin{equation}
\left\{ \mu_{bu},\,  \mu_{bd} \right\} =\left\{ \mu_q-\frac{2}{3} \mu_e -\frac{1}{\sqrt{3}} \mu_8, \,  \mu_q+\frac{1}{3} \mu_e -\frac{1}{\sqrt{3}} \mu_8 \right\}.
\end{equation} 
The rest of the quarks, red up ($ru$), green up ($gu$), red down ($rd$), and green down ($gd$), form the Cooper pairs, which have average chemical potential
\begin{equation}
\mu_C = \frac{\mu_{ru} + \mu_{gd}}{2} = \frac{\mu_{gu} + \mu_{rd}}{2} 
= \mu_q-\frac{1}{6} \mu_e +\frac{1}{2\sqrt{3}} \mu_8 \ .
\end{equation}
The Fermi surfaces of the paired fermion species are locked together
with common Fermi momentum $\mu_C$ \cite{Rajagopal:2000ff}.
The free energy density of 2SC quark matter without strange quarks is
\begin{equation}
\Omega_{2SC} = - \frac{\mu^4_{bu}}{12\pi^2}  - \frac{\mu^4_{bd}}{12\pi^2}  - \frac{\mu^4_{e}}{12\pi^2} -  4 \frac{\mu^4_{C}}{12\pi^2}
- \frac{\mu_C^2 \De^2}{\pi^2} \ .
\end{equation}
The charge neutrality conditions, $\partial \Omega_{\rm{2SC}} /\partial \mu_{e} =0$ and $\partial \Omega_{\rm{2SC}} /\partial \mu_{8} =0$, are satisfied provided $\mu_e =3\left(2+3\cdot 6^{1/3}-6^{2/3} \right) \mu_q/22 $ and $\mu_8= \sqrt{3}\left(12-15\cdot 6^{1/3}+5 \cdot 6^{2/3} \right)  \mu_q/22$,
where we have ignored corrections of order $\De^2/\mu_q^2$.
Thus each chemical potential in the 2SC phase is written in terms of the quark chemical potential $\mu_q$ as
\begin{eqnarray}
\left\{ \mu_{bu}, \mu_{bd}, \mu_{e}, \mu_{C}  \right\} \simeq
\displaystyle
\left\{ 0.566 \mu_q, 1.13 \mu_q , 0.566 \mu_q , 0.934 \mu_q  \right\} \ .
\label{chemical_potential}
\end{eqnarray} 
To obtain a stable 2SC phase we require $\De > \mu_e/2$ 
\cite{Shovkovy:2003uu},
but for  $\mu_e \gtrsim \De > \mu_e/2$ the terms of order 
$\De^2/\mu_q^2$ that we have dropped only modify
\eqn{chemical_potential} by a few percent.
We note that $\mu_{bu} =\mu_{e} = \mu_{bd}/2$ even with the $\Delta^2 /\mu^2_q$ correction.
%We use these values for numerical calculations in this paper.

\subsection{Scattering matrix elements}
\label{sec:scattering}

Our analysis of transport in the 2SC phase of quark matter parallels that of Heiselberg
and Pethick \cite{Heiselberg:1993cr}, who performed perturbative
calculations for unpaired quark matter, assuming the strong coupling
$\al_s$ is small enough to make a perturbative expansion
meaningful. We consider the scattering process of two incoming
particles, particle $1$ of type $i$ and particle $2$ of type $j$,
into two outgoing particles, particle $3$ of type
$i$ and particle $4$ of type $j$.
We denote four-momentum of particle $n$ as $(\epsilon_n, \mathbf{p}_n)$, and $\epsilon_n = |\mathbf{p}_n|$ because the Fermi momenta of the gapless fermions are large enough that we can neglect their masses.
 In the
presence of scattering but no external force, the distribution
function in momentum-position space
of particle $1$, $f_1(\mathbf{x},\mathbf{p}_1,t)$, 
obeys the Boltzmann transport equation
\beq \ba{rcl} \dsp \Bigl(
\frac{\partial}{\partial t} + \mathbf{v}_1 \cdot \nabla_{\mathbf{x}} \Bigr) f_1
&=& \dsp - (2\pi)^4 \sum_{j}\nu_{j} \sum_{234}
|M_{ij}|^2[ f_1 f_2 (1-f_3)(1-f_4)-f_3 f_4 (1-f_1)(1-f_2)] \,
\delta^4(p_{\rm in}-p_{\rm out}) 
\ea
\label{Boltzmann_transport_eq}
\eeq
where $\sum_n= \int d^3p_n/(2\pi)^3$, $ \delta^4 \left(p_{\rm in}-p_{\rm out}\right) = \delta\left(\epsilon_{\rm in}-\epsilon_{\rm out}\right) \delta\left(\mathbf{p}_{\rm in}-\mathbf{p}_{\rm out}\right)$,
and $\nu_j =2$ is the spin factor.
The scattering matrix element $M_{ij}$ is usually decomposed into longitudinal and transverse parts \cite{Weldon:1982aq}, and the longitudinal and transverse components of gauge boson self-energies in the static limit correspond to the Debye mass and the Meissner mass, respectively. According to \cite{Rischke:2003mt,Schmitt:2003aa}, the Meissner mass matrix is diagonal in the rotated $(X,\tilde{Q})$ basis, while the Debye mass matrix is diagonal in the $(T_8,Q)$ basis. Therefore, the two parts of the scattering matrix need to be expressed in the two different bases of the gauge bosons. We now show how to construct the scattering matrix below.  

The indices $i$ and $j$
specify the species of the ungapped fermions, using the basis 
\begin{equation}
\Psi_i = \{\Psi_{bu},\Psi_{bd},\Psi_{e} \}= \{ \mbox{blue up quark ($bu$), blue down quark ($bd$), electron ($e$)} \}.
\end{equation}
The relevant gauge bosons can be written in either the $(T_8,Q)$ or 
the $(X,\tilde{Q})$ basis,
\beq
A_\mu = A^{T_8}_\mu T_8 + A^{Q}_\mu Q 
  = A^{X}_\mu X + A^{\tilde{Q}}_\mu \tilde{Q} \ ,
\eeq
and we write the components as $A^a_\mu$, so $a$ may vary over
$(T_8,Q)$ or  $(X,\tilde{Q})$ depending on the context. The components
are related by
\begin{eqnarray}
\displaystyle A^{X}_\mu &=& 
\cos \varphi \, A^{T_8}_\mu + \sin \varphi \, A^{Q}_\mu 
\\
\displaystyle A^{\tilde{Q}}_\mu &=& -\sin \varphi \, A^{T_8}_\mu + \cos \varphi \, A^{Q}_\mu
\end{eqnarray}
where the mixing angle $\varphi$ is related to the QCD coupling $g$ and the electromagnetic coupling $e$ as \cite{Alford:2010qf}
\begin{eqnarray}
\cos \varphi = \frac{ \sqrt{3} g}{\sqrt{e^2 + 3 g^2}}. 
\end{eqnarray}
We write the covariant derivative as
\begin{equation}
D_\mu \Psi = \Bigl( \partial_\mu - i \sum_{a}A^a_\mu Q^a \Bigr)
\Psi
\end{equation}
where $Q^a$ 
is defined to be the product of the coupling constant and the charge matrix for the ungapped fermions:
\beq
\ba{rcl}
Q^{T_8}
&=&  \dsp
 g \cdot  \mbox{diag} \left(-\frac{1}{\sqrt{3}},
  - \frac{1}{\sqrt{3}},0\right) \\[2ex]
Q^{Q} &=& \dsp e \cdot \mbox{diag} \left(+\frac{2}{3},-\frac{1}{3},-1\right) 
\ea
\eeq
in the $(T_8,Q)$ basis and
\beq
\ba{rcl}
Q^{X}  &=& \dsp g \cos\varphi \cdot \mbox{diag} \left(
  -\frac{1-2 \tan^2\varphi}{\sqrt{3}}, -\frac{1+ \tan^2\varphi}{\sqrt{3}}, 
  -\sqrt{3}\tan^2\varphi\right) \\[2ex]
Q^{\tilde{Q}} &=& \dsp e \cos\varphi \cdot \mbox{diag} \left(1,0,-1\right)
\ea
\eeq
in the $(X,\tilde{Q})$ basis \cite{Alford:2010qf}. 
We will write the $i$th diagonal element as $Q^a_i$,
defined by $(Q^a)_{ij} = Q^a_i \delta_{ij}$ (with no sum over $i$), and we
give the values of $Q^a_i$ in Table~\ref{tab:Qs}.

Because of the screening in a plasma, the gauge bosons acquire self-energies $\Pi_{\mu\nu}$, which then contribute to the gauge field propagator
\beq
\left(D^{ab}_{\mu \nu}\right)^{-1} = g_{\mu \nu} \left(\omega^2-q^2\right) \delta^{a b} + \Pi^{ab}_{\mu \nu} 
\eeq 
where $\omega$ and $\mathbf{q}$ are the energy and momentum transfer. We define $q\equiv|\mathbf{q}|$ and similarly for other momenta. The scattering matrix element for two incoming particles, one with flavor $i$ and four-momentum $\left(\epsilon_1,\mathbf{p}_1 \right)$ and the other with flavor $j$ and four-momentum $\left(\epsilon_2,\mathbf{p}_2 \right)$, is
\begin{eqnarray}
M_{ij} &=& J^\mu_{a,i} \left(D^{ab}_{\mu \nu}\right) J^\nu_{b,j} \\
J^\mu_{a,i}&=& Q^a_i
\bar{u}\left(\mathbf{p}_3\right)\gamma^\mu u\left(\mathbf{p}_1\right)/2p_1 \\
J^\nu_{b,j}&=& Q^b_j \bar{u}\left(\mathbf{p}_4\right)\gamma^\nu u\left(\mathbf{p}_2\right)/2p_2
\end{eqnarray}
where $J^\mu_{a,i}$ and $J^\nu_{b,j}$ are the transition currents,
$\gamma^\mu$ is a Dirac matrix and $u$ is the Dirac spinor. 
%%%%%%%%%%%%%%%%%%%
\begin{table}%[htb] 
\setlength{\tabcolsep}{0.5em}  % desired column spacing
\def\st{\rule[-1.5ex]{0em}{4.5ex}}  % strut to get desired row spacing in table
\begin{tabular}{lcccc}
\hline
\st & $ Q^{T_8} $ & $ Q^{Q}$ & $ Q^{X}$ & $ Q^{\tilde{Q}} \frac{}{}$   \\  \hline
\st blue up ($bu$) & $  - g/ \sqrt{3} $ & $  2 e / 3  $ & $  - g \cos \varphi \left(1- 2 \tan^2 \varphi  \right) /\sqrt{3}  $ & $ e \cos \varphi  \frac{}{}$  
\\ %\hline
\st blue down ($bd$)& $  - g /\sqrt{3} $ & $  -e/3 $ & $  -g \sec\varphi /\sqrt{3} $ & $ 0 \frac{}{}$ 
\\ %\hline
\st electron ($e$) & $  0$ & $  -e $ & $ -\sqrt{3} g \sin \varphi \tan \varphi$ & $ -e \cos\varphi \frac{}{}$ 
\\  %\hline 
\st Cooper pair ($C$) &$ g/ (2 \sqrt{3} )$  & $ e/6$ & $ g \sec \varphi /(2\sqrt{3} ) $ & $ 0 \frac{}{}$  
\\ \hline
\end{tabular}
\caption{Value of $Q^a_i$, the product of the coupling constant and the charge, for each gauge boson $a$ and each gapless fermion $i$. We also show the average for the two quarks in a Cooper pair.
}
\label{tab:Qs}
\end{table}
%%%%%%%%%%%%%%%%%
We split the current into the longitudinal component, $J^l = \mathbf{J} \cdot
\hat{\mathbf{q}} = \omega J^0/q$, and the transverse component, $\mathbf{J}^t = \mathbf{J}- J^l
\hat{ \mathbf{q}}$. We denote the corresponding self-energies in the
propagator as $\Pi^{ab}_l$ and $\Pi^{ab}_t$, respectively. Because the
longitudinal components of the propagator are diagonal in the $(T_8,Q)$ basis
and the transverse components are diagonal in the $(X,\tilde{Q})$ basis
\cite{Rischke:2003mt,Schmitt:2003aa}, we can write the matrix element as

\begin{equation}
M_{ij} 
=
\sum_{a=\{T_8, Q \}} 
\frac{J^0_{a,i} J^0_{a,j}}{q^2 +\Pi^{aa}_l}
- \sum_{a=\{X, \tilde{Q} \}} 
\frac{\mathbf{J}^t_{a,i} \cdot \mathbf{J}^t_{a,j}}{q^2-\omega^2 +\Pi^{aa}_t}
\end{equation}
and after summing over the final spins and averaging over the initial spins, the scattering matrix element can be written as 
\begin{equation}
\ba{rcl}
\left|M_{ij} \right|^2
 &=&  
\displaystyle 
 L_l
\left|   \sum_{a=\{T_8, Q \}} 
\frac{Q^{a}_{i}  \, Q^{a}_{j}  }{q^2+\Pi^{aa}_l} \right|^2
+  L_t \left| \sum_{a=\{X, \tilde{Q} \}} 
 \frac{Q^{a}_{i}  \,  Q^{a}_{j}   }{q^2-\omega^2+\Pi^{aa}_t} 
\right|^2 
\\[5ex]
&&
 \displaystyle
-2  L_{lt} \, \Re \left[
 \left(
\sum_{a=\{T_8, Q \}} 
\frac{Q^{a}_{i}  \,  Q^{a}_{j}    }{q^2+\Pi^{aa}_l}
\right)\!
\left(
\sum_{a=\{X, \tilde{Q} \}} 
\frac{Q^{a}_{i}  \,  Q^{a}_{j}   }{q^2-\omega^2+\Pi^{aa}_t} 
\right)^{\!\!\!*}\, \right] + \delta_{ij} \gamma_{\rm{int}} 
\ea
\label{Mdef}
\end{equation}
where $\gamma_{\rm{int}}$ is the interference term, which is the cross term of two different channels for two identical incoming particles. As discussed in \cite{Shternin:2008es}, however, it is small in the weak-screening approximation, so we neglect $\gamma_{\rm{int}}$. In the limit $\omega \ll p, \mu_q$, $L$'s become
\begin{equation}
\ba{rcl}
L_l
&= & \dsp 
\left(1 - \frac{q^2}{4p^2_1} \right) \left(1 - \frac{q^2}{4p^2_2} \right)
\\[2ex]
L_{lt}
&=& \dsp
\left(1 - \frac{q^2}{4p^2_1} \right)^{\!\!1/2}
\left(1 - \frac{q^2}{4p^2_2} \right)^{\!\!1/2} \cos \theta
\\[2ex]
L_{t}
&=& \dsp
 \left(1 - \frac{q^2}{4p^2_1} \right) \left(1 - \frac{q^2}{4p^2_2} \right) \cos^2 \theta +\frac{q^2}{4 p^2_1} +\frac{q^2}{4 p^2_2} 
\ea
\label{Ldefs}
\end{equation}
where $\theta$ is the angle between $\mathbf{p}_1+ \mathbf{p}_3$ and $\mathbf{p}_2+\mathbf{p}_4$ \cite{Shternin:2008es}.

%Because only the eighth gluon can mediate interactions between the gapless (blue) quarks, the theory is effectively abelian with gauge group $U(1)_{T_8} \times U(1)_Q$, and 

The one-loop correction of self-energies comes from the ungapped fermions and the Cooper pairs.  Following Appendix A in \cite{Alford:2010jf}, we parametrize the longitudinal component $\Pi_l$ and transverse component $\Pi_t$ as
\begin{equation}
\begin{array}{rlr}
\Pi^{aa}_l = 
&
\displaystyle  \sum_{i} \left( q^a_{D,i} \right)^2 \chi_l  + 4  \left( q^a_{D,C} \right)^2 \chi_l 
&  
\mbox{in the $(T_8,Q)$ basis}
\\
\Pi^{aa}_t = 
&
\displaystyle \sum_{i} \left( q^a_{D,i} \right)^2 \chi_t 
+4  \left( q^a_{D,C} \right)^2 \chi_t
+ 4  \left( q^a_{D,C} \right)^2 \chi_{sc} 
& 
 \mbox{in the $(X,\tilde{Q})$ basis}
\end{array}
\label{self_energies_full}
\end{equation}
where $q^a_{D,i}$ and $q^a_{D,C}$ are the Debye masses for a given flavor $i$ and the Cooper pair, respectively, and the factors of 4 in front of
$(q^a_{D,C})^2$ arise from the four different species of Cooper pairs in
the 2SC phase. 
The Debye masses are 
\begin{eqnarray}
\left(q^a_{D,i}  \right)^2
&=&  
\displaystyle
(Q^a_i)^2 \frac{ \mu^2_i}{\pi^2}
\\
\left(q^a_{D,C}  \right)^2
&=&
(Q^a_C)^2 \frac{ \mu^2_C}{\pi^2}
\end{eqnarray}
where $\mu_i$ and $\mu_C$ are the chemical potentials of fermion with
flavor $i$ and the Cooper pair, respectively.
$Q^a_{C}$ is a product of a coupling constant and the
average charge of the two quarks that constitute the pair
\cite{Alford:2005qw} (see Table~\ref{tab:Qs}).
The Cooper pair has $X$ charge but no $\tilde{Q}$ charge, so in the
static limit where $\omega/q \ll1$, $\Pi^{XX}_t$ has a real
component, which gives the Meissner effect, while $\Pi^{\tilde{Q}
  \tilde{Q}}_t$ has an imaginary component, which gives the Landau
damping.  The screening functions, $\chi_l$ and $\chi_t$, are
functions of $\omega$ and $q$, and are calculated in
\cite{Rischke:2000qz,Rischke:2001py,Rischke:2002rz}. In this paper, we use the static
limit of the screening functions \cite{Heiselberg:1993cr,Alford:2005qw},
\begin{equation}
\chi_l = 1, \qquad\chi_t = i \frac{\pi }{4} \frac{\omega}{q},  \qquad
\chi_{sc} = \displaystyle  \frac{1}{3}. 
\end{equation}
Taking the leading order in $\omega/q$, we thus have
\begin{eqnarray}
\Pi^{T_8 T_8}_l 
&=&
\displaystyle  \sum_{i}(Q^{T_8}_i)^2 \, \frac{\mu^2_i}{\pi^2}  + 4 (Q^{T_8}_{C})^2 \frac{\mu^2_{C}}{\pi^2} 
\label{Pi_T8}
\\
\Pi^{ QQ}_l 
&=&
\displaystyle  \sum_{i}(Q^{Q}_i)^2 \, \frac{\mu^2_i}{\pi^2}  + 4 (Q^{Q}_{C})^2 \frac{\mu^2_{C}}{\pi^2} 
\\
\Pi^{XX}_t 
&=&
\displaystyle   \frac{4}{3} (Q^{X}_{C})^2 \frac{\mu^2_{C}}{\pi^2} 
\\
\displaystyle 
\Pi^{ \tilde{Q} \tilde{Q}}_t 
&=&  
\displaystyle 
i \frac{\omega}{q} \Lambda^2 
\qquad \mbox{where}\quad
\Lambda^2 \equiv  \sum_{i}(Q^{\tilde{Q}}_i)^2 \, \frac{\mu^2_i}{4 \pi} 
\label{Pi_Qtilde}
\end{eqnarray}
where $Q$'s are given in Table~\ref{tab:Qs}.

\section{Transport coefficients in a multicomponent system}
\label{Multicomponent}

In preparation for our calculation of the transport properties of the 2SC phase,
we write down transport coefficients in a general multicomponent system at low temperature and high density, $T /\mu \ll 1$, using the linear Boltzmann transport equation in the relaxation time approximation.
We consider an isotropic system which is weakly perturbed from its
equilibrium state. In this case,
the electrical conductivity $\sigma$, the thermal conductivity $\kappa$,
and the shear viscosity $\eta$, are related to the electric current
$j_\alpha$, the heat flux $h_\alpha$, and the shear stress tensor
$\sigma_{\alpha \beta}$, respectively, as \cite{Landau_Vol6}
\begin{eqnarray}
j_\alpha
&=&
- \sigma \partial_\alpha U
\\
h_\alpha
&=& 
- \kappa \partial_\alpha T
\\
\sigma_{\alpha \beta} 
&=& 
- \eta V_{\alpha \beta} 
\end{eqnarray}
where $U$ is the electric potential and $V_{\alpha \beta}$ is the traceless part of the spatial derivative of fluid velocity $\mathbf{V}$, 
\begin{equation}
V_{\alpha \beta} =   \partial_\alpha V_\beta + \partial_\beta V_\alpha -\frac{2}{3} \delta_{\alpha \beta} \mathbf{\nabla} \cdot \mathbf{V} .
\end{equation}
We use $\alpha$, $\beta$, $\lambda$, and $\rho$ as spatial indices.
From kinetic theory, we can write the fluxes on the left-hand
sides as \cite{Lifshitz_Pitaevskii}
\begin{eqnarray}
j_{\alpha}
&=&
\int \frac{d^3 p}{\left( 2\pi \right)^3}  e v_\alpha \,\delta\!f 
\\
h_\alpha
&=&  
\displaystyle \int \frac{d^3 p}{\left( 2\pi \right)^3}  \left(\epsilon -\mu\right) v_\alpha \,\delta\!f 
 \\
\sigma_{\alpha \beta} 
&=& 
\displaystyle \int \frac{d^3 p}{\left( 2\pi \right)^3} p_\alpha v_\beta \,\delta\!f
\end{eqnarray}
where $v_\alpha$ is the particle velocity with $\vert v_\alpha \vert =1$, and
 $\delta\!f$ is a deviation from the equilibrium distribution function $f^0$. 
As we explained in the previous section, the fermions that contribute to the transport properties in the 2SC phase are the blue up quark, blue down quark, and electron. 
In general, we can combine the above equations to write for a multicomponent system as \cite{Flowers:1976,Flowers:1979,Itakura:2007mx}
\begin{equation}
\xi  Y  = \sum_i \nu_i \int \frac{d^3p}{(2\pi)^3} \phi_i \,\delta\!f_i
\label{general_transport}
\end{equation}
where $\nu_i$ is a spin factor for a particle flavor $i$. Instead of writing three different equations for the transport coefficients, we have defined $\xi$ as a transport coefficient for $\sigma$, $\kappa$, or $\eta$ with corresponding macroscopic quantity $Y$ ($- \partial_\alpha U$, $-\partial_\alpha T$, or $-V_{\alpha \beta}$, respectively) and a microscopic quantity
\begin{equation}
\phi_i =  
\left\{ 
\begin{array}{lr}
e_i v_\alpha
& \mbox{Electrical conductivity} 
\\ 
( \epsilon -\mu_i ) v_\alpha
& \mbox{Thermal conductivity} 
\\
p_\alpha v_\beta 
& \mbox{Shear viscosity}
\end{array}\right.
\end{equation}    
respectively. We treat $Y$ and $\phi_i$ as matrices for the shear viscosity and vectors for the thermal and electrical conductivities. 
We write each distribution function with flavor $i$ as
\begin{equation}
f_i = f^0_i + \,\delta\!f_i = \frac{1}{e^{(\epsilon-\mu_i)/T}+1} - \frac{\partial f^0_i}{\partial \epsilon}  \Phi_i
\label{fdefs}
\end{equation}
and we further parametrize the unknown coefficient $\Phi_i$ using the relaxation time approximation
 \cite{Shternin:2007ee,Shternin:2008es}:
\begin{equation}
\Phi_i =  3 \tau_i \psi_i \cdot Y
\end{equation}
where $\tau_i$ is a relaxation time. 
$\psi_i \cdot Y$ denotes the dot product for vectors and the Hadamard product for matrices, i.e., $\psi_i \cdot Y \equiv (\psi_i)_{\alpha \beta} Y^{\alpha \beta}$. 
The numerical factor of $3$ is given so that the definition of the relaxation time agrees with that of Heiselberg and Pethick \cite{Heiselberg:1993cr}.  
$\psi_i$ is a microscopic quantity depending on the transport
phenomena, and the standard forms 
\footnote{We could use the Chapman-Enskog method and write each $\psi_i$ as an infinite sum of trial functions, but one trial function with correct power of momentum is usually sufficient. See, e.g., \cite{Rupak:2007vp,Alford:2009jm}.}
are given as \cite{Lifshitz_Pitaevskii}
\begin{equation}
\psi_i =  
\left\{ 
\begin{array}{lr}
e_i v_\alpha
& \mbox{Electrical conductivity} 
\\ 
\left( \epsilon -\mu_i \right) v_\alpha /T
& \mbox{Thermal conductivity} 
\\
\left( p_\alpha v_\beta -\frac{1}{3}  \delta_{\alpha \beta} \, \mathbf{p} \cdot  \mathbf{v} \right)/2
& \mbox{Shear viscosity}
\end{array}\right.
\end{equation}    
From Eq.~(\ref{general_transport}), we can now define transport coefficient of each component $\xi_i$ as
\begin{equation}
\xi =\sum_i \xi_i =  \xi_{bu}+\xi_{bd}+\xi_{e}
\end{equation}
with
\begin{equation}
\xi_i Y
=
- 3 \tau_i \nu_i  \int \frac{d^3p}{(2\pi)^3} \phi_i \left( \psi_i \cdot Y\right) \frac{\partial f^0_i}{\partial \epsilon}.
\label{transport_Kinetic_1}
\end{equation}
Following the standard procedure, we rewrite $Y$ as
\begin{equation}
Y =  
\left\{ 
\begin{array}{lr}
- \partial_\alpha U =  -\delta^\lambda_{\alpha} \, \partial_\lambda U
& \mbox{Electrical conductivity} 
\\ [1ex]
-\partial_\alpha T =- \displaystyle  \delta^\lambda_{\alpha} \, \partial_\lambda T
& \mbox{Thermal conductivity} 
\\[1ex]
- V_{\alpha \beta} = \displaystyle- \frac{1}{2} \Bigl(\delta^\lambda_{\alpha} \, \delta^\rho_{\beta} \, + \delta^\rho_{\alpha} \, \delta^\lambda_{\beta} \, -\frac{2}{3} \delta_{\alpha \beta} \delta^{\lambda \rho}\Bigr) V_{\lambda \rho} 
&  
\mbox{Shear viscosity}
\end{array}\right.
\end{equation}
and then we can divide the common factor of $Y$ on both sides in Eq.~(\ref{transport_Kinetic_1}) and contract the indices $\alpha$ and $\lambda$ for the electrical and thermal conductivities and the pairs of indices $\alpha$, $\lambda$ and $\beta$, $\rho$ for the shear viscosity. This gives us an expression for a generic transport coefficient
\begin{equation}
\xi_i 
=
- \frac{3 \tau_i \nu_i }{\gamma} \int \frac{d^3p}{(2\pi)^3}\left(  \phi_i \cdot  \psi_i \right) \frac{\partial f^0_i}{\partial \epsilon}
\label{transport_Kinetic_2}
\end{equation}
where $\gamma$ is a numerical factor after contracting the indices: $\gamma=\delta^\alpha_\alpha = 3$ for the electrical and thermal conductivities and $\gamma=\left(\delta^\alpha_\alpha \delta^\beta_\beta  + \delta^\alpha_\alpha- 2 \delta^\alpha_\alpha /3 \right)/2 =5$ for the shear viscosity. 

From the Boltzmann equation, we can obtain another expression for a
transport coefficient. By taking the leading order in the derivative
expansion of the Boltzmann transport equation,
Eq.~(\ref{Boltzmann_transport_eq}), we obtain the linearized Boltzmann
equation:
\begin{eqnarray}
\psi_i \cdot Y  \frac{\partial f^0_1}{\partial \epsilon_1}
&=&
- \frac{(2\pi)^4}{T} \sum_j \nu_j \sum_{234} \left| M_{ij} \right|^2 f^0_1 f^0_2 (1-f^0_3) (1-f^0_4) \,\delta^4(p_{\rm in}-p_{\rm out})
\left( \Phi_1+\Phi_2-\Phi_3-\Phi_4\right).
\label{Linearized_Boltzmann}
\end{eqnarray}
Acting with $ - 3 \tau_i \nu_i \sum_1 \phi_1$ on both sides, we obtain
\begin{eqnarray}
\xi_i Y
=
3\tau_i \frac{(2\pi)^4}{T} \sum_j \nu_i \nu_j \sum_{1234} \left| M_{ij} \right|^2 f^0_1 f^0_2 (1-f^0_3) (1-f^0_4) \delta^4 (p_{\rm in}-p_{\rm out}) 
\phi_1  \left[ 3 \tau_i (\psi_1-\psi_3 ) +3  \tau_j (\psi_2-\psi_4) \right] \cdot Y
\end{eqnarray}
and using the same procedure that led us to  Eq.~(\ref{transport_Kinetic_2}), we have
\begin{eqnarray} 
\xi_i
&=&
\frac{9 \tau_i}{\gamma} \frac{(2\pi)^4}{T} \sum_j \nu_i \nu_j \sum_{1234} \left| M_{ij} \right|^2 f^0_1 f^0_2 (1-f^0_3) (1-f^0_4) \delta^4 (p_{\rm in}-p_{\rm out}) 
\phi_1  \cdot \left[ \tau_i (\psi_1-\psi_3 ) + \tau_j (\psi_2-\psi_4) \right].
\end{eqnarray}
After taking the limit $\omega, T \ll \mu_q$ \cite{Heiselberg:1993cr}, we finally have 
\begin{eqnarray}
\xi_i
&=&
\frac{\tau_i}{\gamma}  \sum_j \nu_i \nu_j \frac{36 T \mu^2_i \mu^2_j }{(2 \pi)^{5}}
\int^\infty_{0} d \omega \left(\frac{\omega/2T}{\sinh(\omega/2T)}\right)^2
 \int^{q_{M}}_{0} dq  \int^{2\pi}_0 \frac{d\theta}{2\pi} 
 \left| M_{ij} \right|^2 
\phi_1  \cdot \left[ \tau_i (\psi_1-\psi_3 ) + \tau_j (\psi_2-\psi_4) \right] 
\label{transport_Boltzmann}
\end{eqnarray}
where $q_M=\mbox{min} \left[ 2p_1,2p_2\right] = \mbox{min} \left[ 2\mu_i,2\mu_j \right]  $ is the maximum momentum
transfer, and $\theta$ is again the angle between $\mathbf{p}_1+
\mathbf{p}_3$ and $\mathbf{p}_2+\mathbf{p}_4$. 
The momentum of an incoming fermion is the Fermi momentum in the limit $T/\mu_q \ll 1$, so we simply replace all $p_1 $ and $p_2$ with $\mu_i$ and $\mu_j$, respectively. 
Equations (\ref{transport_Kinetic_2}) and (\ref{transport_Boltzmann}) can be
used to find the relaxation times
$\tau_i$ for the three gapless fermion species, and thus their
contributions $\xi_i$ to the transport coefficient.

\section{Transport properties in the 2SC phase}
\label{sec:transport}

\subsection{The physics of transport in 2SC quark matter}
\label{sec:qualitative}

Transport in the 2SC phase occurs via the ungapped fermions: the blue up
quark, the blue down quark, and the electron. At a given
temperature, transport is dominated by the
fermion that feels the least influence from the surrounding particles,
since it will have a long relaxation time or mean
free path. The relevant interactions (and their
generators) are as follows. The longitudinal strong interaction ($T_8$)
and the longitudinal electromagnetic interaction ($Q$), which are both
short-ranged because of Debye screening;
the transverse ``rotated'' strong interaction ($X$) which is short-ranged
because of Meissner screening; and the transverse ``rotated'' electromagnetic
interaction ($\tilde{Q}$), which is not screened, only Landau damped.

At low temperatures, where the typical energy transfer $\omega$ is small,
Landau damping (which is proportional to $\omega$ \eqn{Pi_Qtilde}) becomes a
small effect, making the $\tilde{Q}$ interaction long-ranged. The $bu$ quark and
electron, which carry $\tilde{Q}$ charge, therefore experience more scattering
than the $bd$ quark, whose $\tilde{Q}$ charge is zero, so their relaxation time
is short and transport is
dominated by the $bd$ quark. The essential point is that 
at low temperature the long range of the 
$\tilde{Q}$ interaction compensates for its small inherent coupling, 
so particles that feel the $\tilde{Q}$ interaction have suppressed
contributions to transport.
% not sure if we need this:
% However, because the blue up quark feels the strong interaction, it gives 
% shorter relaxation time at intermediate temperature.

At high temperatures, where typical energy transfers are large,
the Landau damping of the $\tilde{Q}$ becomes
more significant, and it no longer has such a long range.
Relaxation times are then determined by the strong interaction
($T_8$ and $X$), so the electron, which has no $T_8$ charge and
only a very small $X$ charge, dominates transport. The next most
important fermion is the blue down quark, simply because its
Fermi momentum is larger \eqn{chemical_potential}, 
so there are more states near its
Fermi surface. 

We therefore expect that as temperature rises, we start off in a regime
dominated by the $bd$ quark, and then make a transition to a regime
dominated by electrons. As we will see, this transition occurs at different
temperatures for different transport properties.

\subsection{Approximation schemes}
\label{sec:approximation}

We now compute the electrical conductivity, the thermal conductivity, and the shear viscosity in the 2SC phase using the formalism developed in Sec.~\ref{Multicomponent}. In each case we perform a numerical calculation and obtain an analytic approximation to it.
The coefficients will be functions of two parameters of microscopic physics, the strong  coupling $\al_s$ and electromagnetic coupling $\al$, and two thermodynamic potentials, the quark chemical potential $\mu_q$ and the temperature $T$.
Following Heiselberg and Pethick \cite{Heiselberg:1993cr}, we will calculate to leading order in $\al_s$ and, for quantities dominated by electromagnetism, in $\al$.
This gives results that are reliable at very high energy scales, but provides at best a rough estimate of the values of the transport coefficients at the energy scales relevant for compact star phenomenology, since the  strong interaction is nonperturbative in that regime. For numerical estimates we will take $\al_s=1$.

The relevant temperature range for compact star phenomenology is from about $10$\,keV to $1$\,MeV while the density regime of interest requires quark chemical potential $\mu_q \sim 400$\,MeV. 
We will therefore make use of an expansion in powers of $T/\mu_q$. 
For numerical computations, we present results for $T/\mu_q$ in the range from $10^{-5}$ to $10^{-3}$.
We can assume that the energy transfer is much smaller than the momentum transfer, $\omega \ll q$, 
because the characteristic energy transfer is of the order of the temperature ($\omega \sim T$), and the characteristic momentum transfer is roughly the screening scale (of order $e\mu$ or $g\mu$) of the relevant gauge bosons.
Terms such as  $q^2- \omega^2 +\Pi^{aa}_t$ in the  transverse component of the scattering matrix element in Eq.~\eqn{Mdef} become $q^2 +\Pi^{aa}_t$.

For physical insight we will also obtain analytic approximations
by using the additional simplifying assumption that the momentum transfer is much smaller than the quark chemical potential,
$q \ll \mu_q$. 
This is a good approximation for the transverse component of the $\tilde{Q}$ interaction \cite{Shternin:2008es}, because the self-energy of the $\tilde{Q}$ boson is Landau-damped, and the characteristic momentum transfer is $q \sim (e^{2} \mu^{2}_q  T)^{1/3} \ll \mu_q$. Therefore, even for the numerical computations, we use the analytical expression for the $\tilde{Q}$ interaction (third lines of Eqs.~\eqn{integral_identity_q} and \eqn{integral_identity_omega}) by taking the limit $q \ll \mu_q$ in Eq.~\eqn{transport_Boltzmann}. 
As pointed out by Shternin and Yakovlev \cite{Shternin:2008es}, the approximation is not always reliable for screened interactions: high momentum transfer processes sometimes play an important role.
However, we will show that these analytic results agree well with the numerical results in the 2SC phase, 
and they provide us with a physical understanding of the numerical results.

%We always ignore the next leading order of $\omega/q$ because the integral of Eq.~(\ref{transport_Boltzmann}) is typically peaked at $q \sim \sqrt{\Pi^{aa}_{l,t}}$, and we always have $\omega \sim T \ll \sqrt{\Pi^{aa}_{l,t}}$ for the parameters we choose.  

\subsection{Electrical conductivity of $\tilde{Q}$ charge}
\label{sec:electrical_conductivity}

\begin{figure}
\includegraphics[width=4.5in]{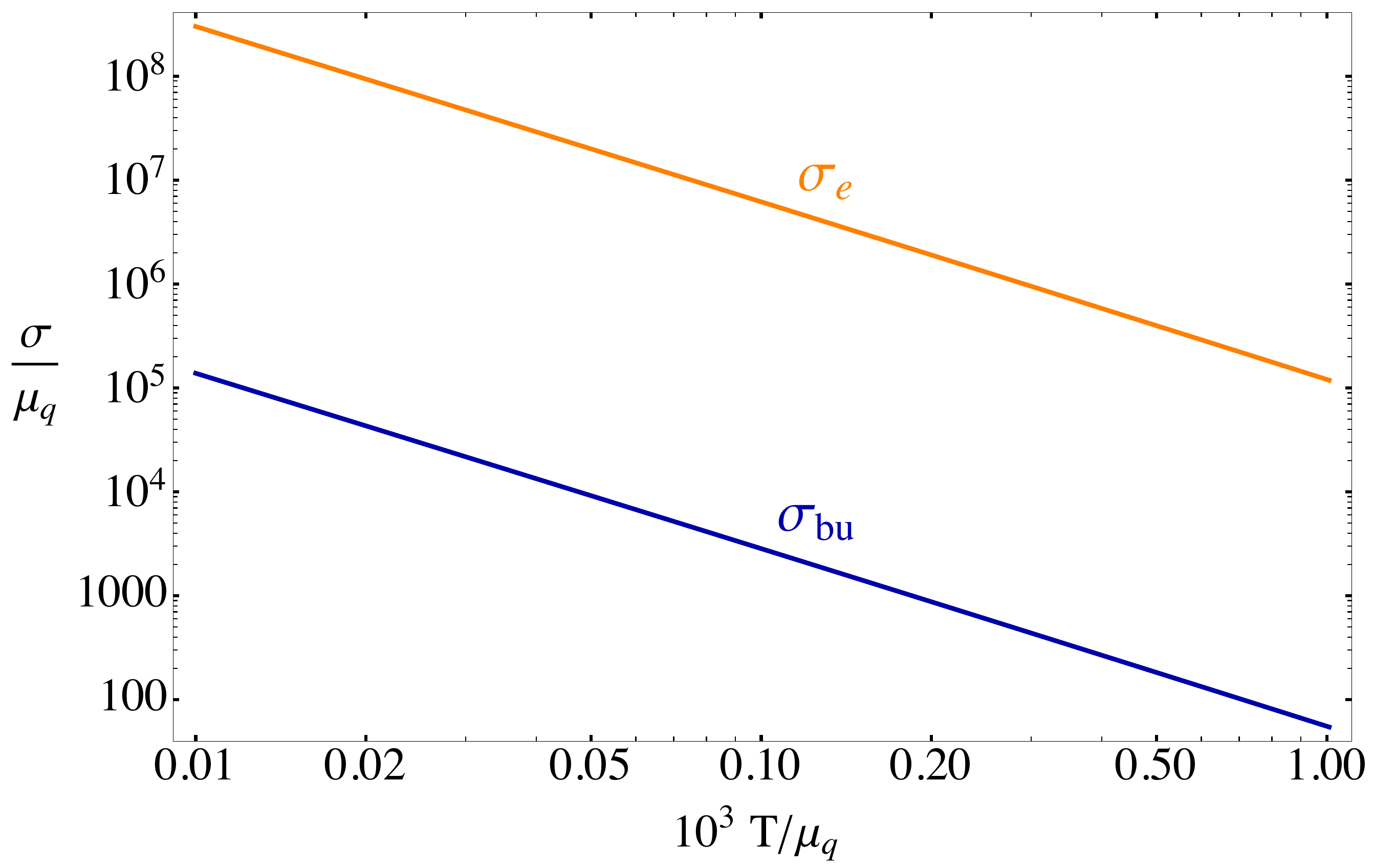}
\caption{\label{2SC_ElectricalConductivity}
(Color online)
Numerically calculated electrical ($\tilde{Q}$) conductivity as
a function of temperature, both expressed in units of the
quark chemical potential $\mu_q$, taking strong interaction coupling
$\al_s=1$. The electrons dominate because the $bu$ relaxation time
is shortened by its strong interaction with the $bd$ quarks.}
\end{figure}

In the 2SC phase, electrical conductivity involves
$\tilde{Q}$ charge rather than $Q$ charge. A charged current
produces magnetic fields, but $X$ magnetic fields are Meissner screened,
so only the $\tilde{Q}$ current exists in the bulk of 2SC matter. 
The expected behavior discussed in Sec.~\ref{sec:qualitative} is
affected by the fact that $bd$ quarks have no $\tilde{Q}$ charge, so
their low-temperature dominance of transport is not relevant to
electrical conductivity.
$\tilde{Q}$ charge is carried by the electron and the blue up quark,
which have  $\tilde{Q}$ interactions with each other. However,
the blue up quark has a shorter relaxation time because of its 
additional strong interactions with the blue down quarks, so
conductivity will be dominated by the electron.
We will see below how our calculations confirm this expectation.

For the electrical conductivity of $\tilde{Q}$ charge, we have $\phi_i =
Q^{\tilde{Q}}_{i} v_\alpha$  and $\gamma=3$ in
Eq.~(\ref{transport_Kinetic_2}), which then gives the Drude result
for the conductivity of species $i$ with relaxation time $\tau^\sigma_i$,
\begin{equation}
\sigma_i = \tau^\sigma_{i} \frac{\mu^2_i ( Q^{\tilde{Q}}_{i})^2}{\pi^2}.
\label{sigma_kinetic}
\end{equation}
Because $Q^{\tilde{Q}}_{bd}$ is zero,  $\sigma_{bd}=0$.
To calculate $\tau^\sigma_{i}$ using Eq.~(\ref{transport_Boltzmann}), we write
\begin{equation}
\phi_1\cdot \left[ \tau^\sigma_i \left( \psi_1-\psi_3 \right)+\tau^\sigma_j \left( \psi_2-\psi_4\right) \right]
=
\frac{ ( Q^{\tilde{Q}}_{i})^2}{2p^2_1}\left( \tau^\sigma_i  -\tau^\sigma_j  \frac{p_1 Q^{\tilde{Q}}_{j}}{ p_2 Q^{\tilde{Q}}_{i}} \right)  q^2 
\end{equation}
where we have ignored the terms suppressed by factors of $\omega / p_{1,2}$.
Using the above two equations in Eq.~(\ref{transport_Boltzmann}), we find
\begin{equation}
1= \frac{3T}{4 \pi^3} \sum_j \frac{\mu^2_j}{ \mu^2_i} \left( \tau^\sigma_i   -\tau^\sigma_j  \frac{\mu_i Q^{\tilde{Q}}_{j}}{\mu_j Q^{\tilde{Q}}_{i}} \right)  s^{\sigma}_{ij}
\label{master_electrical}
\end{equation}
for a flavor $i=bu$ or $e$, where
\begin{equation}
s^\sigma_{ij} 
=
\int^\infty_{0} d\omega \left(\frac{\omega/2T}{\sinh(\omega/2T)}\right)^2
\int^{q_M}_{0} dq  \int^{2\pi}_0 \frac{d\theta}{2\pi} 
\left| M_{ij} \right|^2 q^2
\label{s_sigma}
\end{equation}
which is symmetric in exchanging $i$ and $j$.
We can then solve Eq.~\eqn{master_electrical} for the relaxation times, $\tau^\sigma_{bu}$ and $\tau^\sigma_{e}$, and find
\begin{eqnarray}
\frac{1}{\tau^{\sigma}_{i}}
=
\frac{3 T}{4 \pi^3} s^\sigma_{i,bd}\, s^\sigma_{bu,e} \left(\frac{1}{s^\sigma_{bu,bd}} + \frac{4}{s^\sigma_{bu,e}} + \frac{1}{s^\sigma_{bd,e}} \right)
\label{sigma_tau}
\end{eqnarray}
where $i=bu$ or $e$. 
Even though $\sigma_{bd}=0$, the blue down quark has some effect on the conductivity because its interactions with the charged fermions,
particularly the $bu$ quark, affect their relaxation rates.
We numerically integrate Eq.~(\ref{s_sigma}) 
using the scattering matrix element $M_{ij}$  given in Eq.~\eqn{Mdef}, with the charges from Table~\ref{tab:Qs} with $\alpha=1/137$ and $\alpha_s=1$, $L$'s from Eq.~\eqn{Ldefs} with $p_1 = \mu_i$ and $p_2 =\mu_j$, the boson self-energies from Eqs.~\eqn{Pi_T8} to \eqn{Pi_Qtilde}, and the chemical potentials from Eq.~(\ref{chemical_potential}). $s^\sigma$ is then a dimensionless function of $T/\mu_q$.
We use the numerical value of $s^\sigma$ in Eq.~(\ref{sigma_tau}) to obtain $\tau^\sigma_i \mu_q$, which then gives $\sigma_i /\mu_q$ by Eq.~(\ref{sigma_kinetic}). The results are plotted in Fig.~\ref{2SC_ElectricalConductivity},
and the best fits are $\sigma_{bu} /  \mu_q = 0.000672/((T / \mu_q)^{5/3}+2.11 \, (T / \mu_q)^2)$ and $\sigma_e /\mu_q = 1.46/((T / \mu_q)^{5/3}+2.11 \,(T / \mu_q)^2)$. We note these fits can be extrapolated to arbitrary low temperature.

We see in Fig.~\ref{2SC_ElectricalConductivity} that the electrons
dominate the conductivity, obeying a $T^{-5/3}$ power law.
To understand this we now derive an approximate
analytic expression for the electrical
conductivity by assuming that
the momentum transfer $q$ is much less than the
typical Fermi momenta. 
In this limit, $q \ll \mu_q \sim p_{1,2}$, Eq.~\eqn{Ldefs} simply becomes $L_l = 1$, $L_{lt} = \cos \theta$, and $L_t = \cos^2 \theta$. 
Furthermore, the upper interval of the $q$ integral in Eq.~(\ref{s_sigma}) 
may be taken as infinity because the integrand is very small for
$q\gtrsim q_M$. The integral in Eq.~(\ref{s_sigma}) can then be
performed analytically \cite{Heiselberg:1993cr} using the identity
\begin{equation}
\int^\infty_{0} d\omega \left(\frac{\omega/2T}{\sinh(\omega/2T)}\right)^2
\int^{\infty}_{0} dq
\, \mbox{Re} \left[  \frac{ q^2 }{\left(q^2 + \Pi^{aa}\right)\left(q^2 + \Pi^{bb} \right)^* } \right]
  = 
 \left\{ 
\begin{array}{lr}
\displaystyle \frac{\pi^3 \, T }{6 \left(\sqrt{\Pi^{aa}} + \sqrt{\Pi^{bb}} \right)}
& 
\mbox{if neither $a$ nor $b$ is $\tilde{Q}$}
\\[4ex]
\displaystyle\frac{ \pi^3 \, T }{  6 \sqrt{\Pi^{aa}}  } +\mathcal{O} \left(\frac{T^{4/3}}{\mu^{4/3}_q}\right)
& 
\mbox{if $b$ is $\tilde{Q}$ and $a$ is not}
\\[4ex]
\displaystyle \frac{\pi \Gamma (8/3)  \zeta (5/3) \, T^{2/3} }{ 3 \Lambda^{2/3}  }
& 
\mbox{if $a$ and $b$ are both $\tilde{Q}$} 
\end{array} \right.
\label{integral_identity_q}
\end{equation}
where $\Pi^{aa}$ is a self-energy of a boson of type $a$,
with the convention that $\Pi^{aa}=\Pi^{aa}_l$ if $a$ is either $T_8$ or $Q$ and $\Pi^{aa} = \Pi^{aa}_{t}$ if $a$ is either $X$ or $\tilde{Q}$, as given in Eqs.~\eqn{Pi_T8} to \eqn{Pi_Qtilde}. 
For the first case, both the gauge boson self-energies 
$\Pi^{aa}$ and $\Pi^{bb}$ are independent of $\omega$ and $q$, 
so the integration can be performed straightforwardly. 
In the second case, one of the self-energies
is for the $\tilde Q$ photon, which is dominated by Landau damping,
$\Pi^{\tilde{Q} \tilde{Q}}_t = i \omega \Lambda^2 /q$ \eqn{Pi_Qtilde}.
In this case, we first perform the $q$ integral exactly, then keep the leading order in $\omega/ \mu_q$ because $\omega \sim T  \ll \mu_q$, and finally perform the $\omega$ integral exactly. This is equivalent to doing the integral exactly by neglecting $\Pi^{\tilde{Q} \tilde{Q}}_t$ because $\omega /q  \ll 1$.
In the third case, however, both the self-energies are for
$\tilde{Q}$ photons; their Landau damping acts as the regulator
of an infrared divergence and the integral
scales as $(T/\Lambda)^{2/3}$. We call this the ``$\tilde{Q}$-interaction
term."
We thus find 
\begin{eqnarray}
s^\sigma_{ij} 
&=&
\frac{\pi^3  T}{6}
\sum_{a,b=\{ T_8,Q \}} 
\frac{Q^a_{i} \,  Q^a_{j} \, Q^b_{i} \, Q^b_{j}}{\sqrt{\Pi^{aa}_l}+\sqrt{\Pi^{bb}_l}}
+ \frac{\pi^3 T}{12} 
\left( \frac{( Q^X_{i})^2 ( Q^X_{j})^2}{ 2 \sqrt{\Pi^{XX}_t}} 
+\frac{2  Q^X_{i} \, Q^X_{j} \, Q^{\tilde{Q}}_{i} \, Q^{\tilde{Q}}_{j} }{\sqrt{\Pi^{XX}_t}}
\right)
\nonumber
\\
&&
\displaystyle
+ \frac{\pi \Gamma(8/3) \zeta(5/3)T^{2/3} }{6}  \frac{( Q^{\tilde{Q}}_{i})^2 ( Q^{\tilde{Q}}_{j})^2}{\Lambda^{2/3}} 
\label{s_electrical_conductivity}
\end{eqnarray}
where the self-energies are given in Eqs.~\eqn{Pi_T8} to \eqn{Pi_Qtilde}. The first term comes from the longitudinal interactions and the rest comes from the transverse interactions. 
As shown in \cite{Heiselberg:1993cr}, the $\tilde{Q}$-interaction term has 
a lower power of the temperature
because Landau damping gives a small contribution to the self-energy at low energy transfer, making the
 $\tilde{Q}$ interaction long-ranged at low temperature.
If we use this analytic approximation in Eqs.~\eqn{sigma_tau} and \eqn{sigma_kinetic}, then we obtain values of $\tau^\sigma_{bu}$ and $\tau^\sigma_{e}$ 
that agree with the numerical evaluation to within 35\% and 4\%, respectively, at temperatures up to $10^{-3} \mu_q$. 

We can explain qualitatively why $\sigma_e$ is much larger than $\sigma_{bu}$. 
From Eqs.~\eqn{sigma_kinetic} and \eqn{sigma_tau}, we have $\sigma_{e}/\sigma_{bu} = \tau_{e}/\tau_{bu} = s^\sigma_{bu,bd}/s^\sigma_{e,bd}$
where the first equality comes from the fact that the blue up quark and the electron have the same chemical potential and the opposite $\tilde{Q}$ charge. 
From Eq.~\eqn{s_electrical_conductivity}, we find that
$s^\sigma_{bu,bd}$ is proportional to $g^3$ because the main interactions are the screened strong interactions by the $T_8$ and $X$ bosons. 
$s^\sigma_{bd,e}$ is proportional to $e^3$ because it is dominated by the longitudinal screened electromagnetic $Q$ interaction.
Thus we can estimate that $\sigma_{e}/\sigma_{bu} \sim (g/e)^3 \sim 10^{3}$, which qualitatively agrees with Fig.~\ref{2SC_ElectricalConductivity}.

We now explicitly show the analytical expression for the dominant contribution of $\sigma_e$.
The first term in Eq.~\eqn{sigma_tau} is negligible compared to the other two terms because $s^\sigma_{bu,bd}$ is proportional to a scattering amplitude of the strong interaction. 
Thus we have
\begin{equation}
\frac{1}{\tau^\sigma_e} 
=
\frac{3T}{4 \pi^3} \left(4 s^\sigma_{bd,e} + s^\sigma_{bu,e} \right).
\end{equation}
The leading term of $s^\sigma_{bu,e}$ is the $\tilde{Q}$ interaction term, which is proportional to $(T/ \mu_q)^{2/3}$ from Eq.~\eqn{integral_identity_q} (third case), while the leading term of $s_{bd,e}$ is $(T/\mu_q)$ from \eqn{integral_identity_q} (first and second cases).
If we only keep the $\tilde{Q}$ interaction term in $s^\sigma_{bu,e}$, then we have
\begin{equation}
\ba{rcl}
\sigma_e
&= & 
\displaystyle
\frac{ \mu^2_e  (Q^{\tilde{Q}}_{e})^2 }{\pi^2}  \frac{4 \pi^3}{3T s^\sigma_{bu,e} }
=
\dsp
\frac{ \mu^2_e  (Q^{\tilde{Q}}_{e})^2 }{\pi^2}  \frac{8 \pi^2 \Lambda^{2/3}}{ \Gamma (8/3) \zeta (5/3)T^{5/3} (Q^{\tilde{Q}}_{bu})^2 (Q^{\tilde{Q}}_{e})^2} 
\\[3ex]
&\simeq& \dsp
\frac{\mu^2_{q} e^2}{\pi^2} \frac{0.0433}{\alpha^{5/3} T (T/\mu_q)^{2/3}} 
\ea 
\label{electrical_conductivity}
\end{equation}
which agrees with the numerical result to 5\% at $T =  10^{-5}\mu_q$ and 22 \% at $T =  10^{-3}\mu_q$.
We thus conclude that for the electrical conductivity of $\tilde{Q}$ charge, the dominant contribution comes from the electron, and the relevant scattering process in leading order of $T/\mu_q$ is between the electron and the blue up quark via the $\tilde{Q}$ interaction.

\subsection{Thermal conductivity}
\label{sec:thermal_conductivity}

\begin{figure}
\includegraphics[width=4.5in]{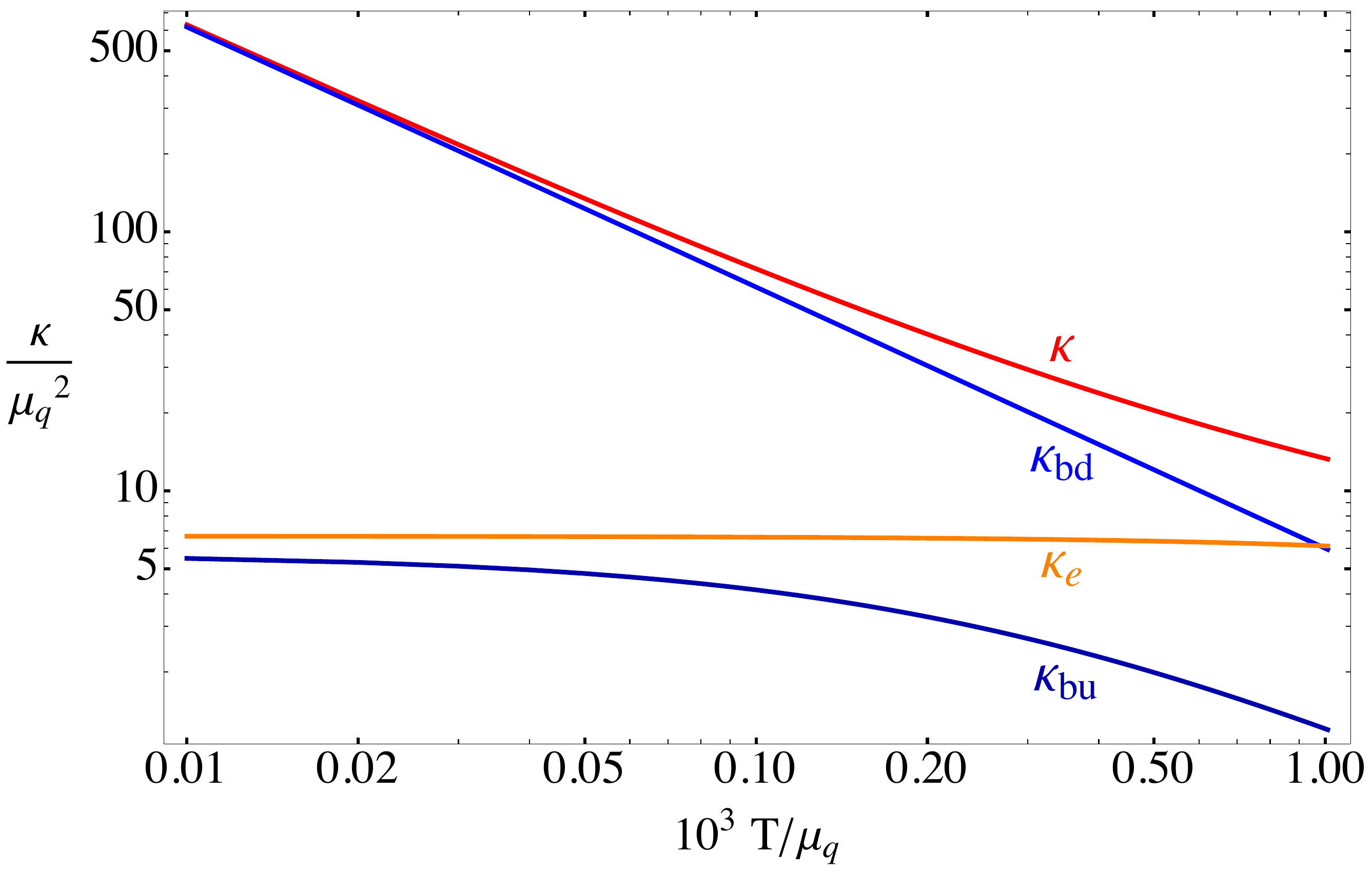}
\caption{\label{2SC_ThermalConductivity}
(Color online)
Numerically calculated thermal conductivity in 
units of quark chemical potential $\mu_q$ in the 2SC phase with $\alpha_s=1$. In this temperature range we see the crossover from
electron domination at high temperature to blue down quark domination
at low temperature (see Sec.~\ref{sec:qualitative}). }
\end{figure}

For the thermal conductivity, the discussion of Sec.~\ref{sec:qualitative}
applies straightforwardly, as we now demonstrate.
We have $\phi_i = (\epsilon-\mu_i) v_\alpha $ and $\gamma=3$ in Eq.~(\ref{transport_Kinetic_2}), which then gives
\begin{equation}
\kappa_i = \tau^{\kappa}_{i} \frac{\mu^2_i T}{3}.
\label{thermal_kinetic}
\end{equation}
To calculate $\kappa_i$ using Eq.~(\ref{transport_Boltzmann}) in the limit $\omega \ll q$ and $T \ll \mu_q$, we write
\begin{eqnarray}
 \phi_1\cdot  \left[ \tau^{\kappa}_i  \left( \psi_1-\psi_3 \right)+\tau^{\kappa}_j \left( \psi_2-\psi_4\right) \right]
&=&
 \frac{\phi_1 - \phi_3 }{2T} \cdot \left[ \tau^{\kappa}_i \left( \psi_1-\psi_3 \right)+\tau^{\kappa}_j \left( \psi_2-\psi_4\right) \right]
 \nonumber
 \\
&=&
\frac{\omega^2}{2T} \left[ \tau^{\kappa}_i  +  \frac{q^2}{4 p_1 p_2} \tau^{\kappa}_j 
-\cos \theta  \sqrt{\left(1- \frac{q^2}{4p^2_1} \right)  \left(1- \frac{q^2}{4p^2_2} \right)}
\tau^{\kappa}_j \right] .
\end{eqnarray}
Using the above two equations in Eq.~(\ref{transport_Boltzmann}), we have
\begin{equation}
1= \frac{9}{4 \pi^5 T} \sum_j \mu^2_j \left( \tau^{\kappa}_i  s_{ij} + \tau^{\kappa}_j \tilde{s}_{ij}\right)
\label{thermal_master}
\end{equation}
for each flavor $i$, where
\begin{equation}
\ba{rcl}
s^{\kappa}_{ij} 
&=& \dsp
\int^\infty_{0} d\omega \left(\frac{\omega/2T}{\sinh(\omega/2T)}\right)^2
 \int^{q_M}_{0} dq  \int^{2\pi}_0 \frac{d\theta}{2\pi} 
 \left| M_{ij} \right|^2 \omega^2 
 \\[3ex]
 \tilde{s}^{\kappa}_{ij}
 &=& \dsp
\int^\infty_{0} d\omega \left(\frac{\omega/2T}{\sinh(\omega/2T)}\right)^2
 \int^{q_M}_{0} dq  \int^{2\pi}_0 \frac{d\theta}{2\pi} 
 \left| M_{ij} \right|^2 \omega^2 
\left(  \frac{q^2}{4 p_1 p_2} -\cos \theta  \sqrt{\left(1- \frac{q^2}{4p^2_1} \right)  \left(1- \frac{q^2}{4p^2_2} \right)}  \right)
\ea
\label{s-kappa}
\end{equation}
which are symmetric in exchanging $i$ and $j$.   
We numerically evaluate both $s^{\kappa}_{ij}$ and $\tilde{s}^{\kappa}_{ij}$ as functions of $T/\mu_q$ using the scattering matrix element $M_{ij}$  given in Eq.~\eqn{Mdef}, with the charges from Table~\ref{tab:Qs} with $\alpha=1/137$ and $\alpha_s=1$, $L$'s from Eq.~\eqn{Ldefs} with $p_1 = \mu_i$ and $p_2 =\mu_j$, the boson self-energies from Eqs.~\eqn{Pi_T8} to \eqn{Pi_Qtilde}, and the chemical potentials from Eq.~(\ref{chemical_potential}). 
We then solve the three relaxation times from the three equations in Eq.~(\ref{thermal_master}), and we use $ \tau^\kappa_i \mu_q$ in Eq.~(\ref{thermal_kinetic}) to obtain $\kappa_i/\mu^2_q$. 
The results are plotted in Fig.~\ref{2SC_ThermalConductivity},
and the best fits are $\kappa_{bu} /\mu^2_q= 5.69 /(1+3720 \, (T/\mu_q))$, $\kappa_{bd} / \mu^2_q=0.00617/(T/ \mu_q)$, and $\kappa_e /\mu^2_q = 6.70/(1+6.92 \, (T /\mu_q)^{2/3}) $. We note these fits can be extrapolated to arbitrary low temperature.

We see in Fig.~\ref{2SC_ThermalConductivity} the expected gradual transition from a low-temperature regime
dominated by blue down quarks to a high-temperature regime dominated
by electrons. 
We now derive approximate analytic expressions to 
account for this behavior,
by assuming that the momentum transfer $q$ is much less than the
typical Fermi momenta \cite{Heiselberg:1993cr}.
We use the identities
\begin{eqnarray}
&&
\int^\infty_{0} d\omega \left(\frac{\omega/2T}{\sinh(\omega/2T)}\right)^2
\int^{\infty}_{0} dq
\, \mbox{Re}
\left[  \frac{ \omega^2 }{\left(q^2 + \Pi^{aa} \right)\left(q^2 + \Pi^{bb} \right)^* } \right] 
\nonumber
 \\ 
&&  = \left\{ 
\begin{array}{lr}
\displaystyle  \frac{2\pi^5 \, T^3}{15 \left(\Pi^{aa} \sqrt{\Pi^{bb}} +\Pi^{bb} \sqrt{\Pi^{aa}} \right)}
& 
\mbox{if neither $a$ nor $b$ is $\tilde{Q}$}
\\ 
\displaystyle \frac{\pi \Gamma(14/3) \zeta(11/3) \, T^{8/3}}{3 \Pi^{aa}  \Lambda^{2/3}} 
- 
\frac{2\pi^5 \, T^3}{15 \left( \Pi^{aa} \right)^{3/2}}
+ 
\mathcal{O} \left(   \frac{T^{10/3}}{ \mu^{10/3}_q } \right)
& 
\mbox{if $b$ is $\tilde{Q}$ and $a$ is not}
\\ 
\displaystyle  \frac{\pi  \zeta (3) \, T^2}{\Lambda^2}
& 
\mbox{if $a$ and $b$ are both $\tilde{Q}$} 
\end{array} \right.
\label{integral_identity_omega}
\end{eqnarray}
For the second case, we keep up to $T^3/\mu^3_q $, which is the same order as the first case.
This allows us to obtain a closed form for $s^\kappa$ and $\tilde{s}^\kappa$ up to $T^3/ \mu^3_q$
\begin{eqnarray}
s^\kappa_{ij} 
&=&
\frac{\pi \zeta(3) T^2}{2} \frac{\left( Q^{\tilde{Q}}_{i} \right)^2 \left( Q^{\tilde{Q}}_{j} \right)^2}{\Lambda^2} 
+ \frac{\pi \Gamma(14/3) \zeta(11/3) T^{8/3}}{3} \frac{Q^X_i \, Q^X_j  \,  Q^{\tilde{Q}}_i \, Q^{\tilde{Q}}_{j} }{ \Pi^{XX}_t \Lambda^{2/3}}
\nonumber
\\
&&
+\frac{2 \pi^5  T^3}{15} \left[ 
-  \frac{Q^X_i \, Q^X_j  \,  Q^{\tilde{Q}}_i \, Q^{\tilde{Q}}_{j}}{\left(\Pi^{XX}_t \right)^{3/2}}
+
\sum_{a,b=\left\{ T_8,Q\right\}} 
\frac{Q^a_i \, Q^a_j  \,  Q^{b}_i \, Q^{b}_{j} }{\Pi^{aa}_l\sqrt{\Pi^{bb}_l}+\Pi^{bb}_l\sqrt{\Pi^{aa}_l}}
+\frac{1}{2}
\frac{\left( Q^X_{i} \right)^2 \left( Q^X_{j} \right)^2}{2 \, \Pi^{XX}_t \sqrt{\Pi^{XX}_t}} 
\right]
\label{s_kappa}
\end{eqnarray}
and 
\begin{eqnarray}
\tilde{s}^\kappa_{ij} 
&=&
\frac{\pi \Gamma(14/3) \zeta(11/3) T^{8/3}}{3} 
\sum_{a=\left\{ T_8,Q\right\}} 
\frac{ Q^a_{i} \, Q^a_{j} \, Q^{\tilde{Q}}_{i} \, Q^{\tilde{Q}}_{j} }{\Pi^{aa}_l \Lambda^{2/3}}
\nonumber
\\
&&
+ \frac{2 \pi^5 T^3}{15} \sum_{a=\left\{ T_8,Q\right\}} 
\left[
- \frac{Q^a_{i} \, Q^a_{j} \, Q^{\tilde{Q}}_{i} \, Q^{\tilde{Q}}_{j} } {\left(\Pi^{aa}_t \right)^{3/2}}
+\frac{ Q^a_{i} \, Q^a_{j} \, Q^{X}_{i} \, Q^{X}_{j} }{\Pi^{aa}_l\sqrt{\Pi^{XX}_t}+\Pi^{XX}_t\sqrt{\Pi^{aa}_l}}
\right] .
\label{st_kappa}
\end{eqnarray}
The term proportional to $T^2$ in $s^\kappa$ comes from the $\tilde{Q}$ interaction, while the terms proportional to $T^3$ in $s^{\kappa}$ and $\tilde{s}^{\kappa}$ come from the screened interactions by the $Q$, $T_8$, and $X$ bosons. The term proportional to $T^{8/3}$ is the cross term of the $\tilde{Q}$ interaction and screened interactions in the scattering matrix $|M_{ij}|^2$. As shown in \cite{Heiselberg:1993cr}, the $\tilde{Q}$-interaction term has a lower power of temperature because of the Landau damping, and it is the leading term at lower temperature. 
Using these expressions in Eq.~(\ref{thermal_master}), we
can solve for the $\tau^\kappa_i$ in closed form. These 
expressions are, however, lengthy, so we only show results for $\alpha_s=1$ and $\alpha=1/137$, expanding $1/\tau^\kappa_i$ to order of $T/\mu_q$ to obtain
\begin{eqnarray}
\kappa_{bu} 
&\simeq &
\displaystyle
\frac{\mu^2_{q} }{3} \frac{16.9}{1  +75.3 (T/\mu_q)^{2/3}+3350 (T/\mu_q)} 
\label{kappa_bu}
\\
\kappa_{bd} 
&\simeq &
\displaystyle
\frac{\mu^2_{q}}{3} \frac{0.0189}{ T/\mu_q} 
\label{kappa_bd}
\\
\kappa_{e} 
&\simeq &
\displaystyle
\frac{\mu^2_{q} }{3} \frac{20.0}{1+29.9 (T/\mu_q)^{2/3}-58.8 (T/\mu_q)} 
\label{kappa_e}
\end{eqnarray}
which agree with the numerical results
to within $11\%$ for $\kappa_{bu}$, $6\%$
for $\kappa_{bd}$, and $12\%$ for $\kappa_e$
at temperatures up to $10^{-3}\mu_q$.
In the denominators of the expressions above for $\ka_i$, we see terms proportional to $T^0$, $T^{2/3}$, and $T$. The terms of order $T^0$
arise from electromagnetic scattering of the relevant fermion by the
background of charged gapless fermions. This interaction is mediated
by $\tilde{Q}$ photons, and the power of $T$ is determined by the Landau damping of the transverse $\tilde{Q}$ photon propagator as shown in Eq.~\eqn{Pi_Qtilde}. 
The terms of order $T^{2/3}$ arise from the cross term of $\tilde{Q}$ and screened interactions in $s^{\kappa}$ and $\tilde{s}^\kappa$, and the terms of order $T$ arise from the screened interactions in $s^{\kappa}$ and $\tilde{s}^\kappa$.
We can now use these physical insights to obtain analytic expressions
for some of the numerical coefficients in \eqn{kappa_bu}, \eqn{kappa_bd}, and \eqn{kappa_e}.

If we take the $\tilde{Q}$-interaction term and ignore the other terms in $s^\kappa$ and $\tilde{s}^\kappa$, then we can solve Eq.~\eqn{thermal_master} for $\tau^\kappa_{bu}$ and $\tau^\kappa_{e}$ exactly.
For $\kappa_e$, from Eq.~\eqn{thermal_kinetic}, we have
\begin{equation}
\kappa_{e} 
=
\frac{\mu^2_{e} T}{3} \frac{4\pi^5 T}{9 \mu^2_{e}
  (s^\kappa_{e,e}+s^\kappa_{e,bu})} 
=
  \frac{4 \pi^4}{27 \zeta(3)} \frac{\Lambda^2}{\bigl( Q^{\tilde{Q}}_e \bigr)^4 }
\simeq
\frac{\mu^2_q }{3}  \frac{0.146 }{\alpha } 
\label{kappa_e_leading}
\end{equation}
which gives the leading $T^0$ term in Eq.~\eqn{kappa_e}.
For $\kappa_{bu}$, the same approximation yields an expression similar to Eq.~\eqn{kappa_e_leading}, but this is not a good approximation at $T/\mu_q \simeq 10^{-3}$ because the 
coefficients of the higher-order terms of $T/\mu_q$ are large as we can see in Eq.~\eqn{kappa_bu}. This is because they arise from the strong interaction. 
For $\kappa_{bd}$, 
the leading order for the denominator is $T$, which comes from the screened $T^8$, $X$, and $Q$ interactions, but we can ignore the $Q$ interaction because $e/g$ is small. 
The analytic solution then becomes
\begin{eqnarray}
\kappa_{bd}  = \frac{\mu^2_{bd} T}{3} \frac{16\pi^5 T}{9 \mu^2_{bd} (s^\kappa_{bu,bd}+4 s^\kappa_{bd,bd}+4 \tilde{s}^\kappa_{bd,bd})} 
\label{kappa_bd_analytic}
\end{eqnarray}
with
\begin{eqnarray}
s^\kappa_{bu,bd}+4 s^\kappa_{bd,bd}+4\tilde{s}^\kappa_{bd,bd} 
&=& 
\frac{2 \pi^5  T^3}{15} \left[ 
\frac{\bigl( Q^{T_8}_{bu} \bigr)^2 \bigl( Q^{T_8}_{bd} \bigr)^2+4\bigl( Q^{T_8}_{bd} \bigr)^4}{2\bigl(\Pi^{T_8 T_8}_l \bigr)^{3/2}}
+
\frac{\bigl( Q^X_{bu} \bigr)^2\bigl( Q^X_{bd} \bigr)^2+4\bigl( Q^X_{bd} \bigr)^4}{4 \, \bigl(\Pi^{XX}_t\bigr)^{3/2}} 
\right.
\nonumber
\\
&&
\left.
+
\frac{4 \bigl( Q^{T_8}_{bd} \bigr)^2 \bigl( Q^X_{bd} \bigr)^2}{\Pi^{T_8 T_8}_l\sqrt{\Pi^{XX}_t}+\Pi^{XX}_t\sqrt{\Pi^{T_8 T_8}_l}}
\right]
\nonumber
\end{eqnarray}
which gives Eq.~\eqn{kappa_bd}. 
Therefore, the relevant scattering process in leading order of $T/\mu_q$ for $\kappa_e$ is between electrons and blue up quarks via the $\tilde{Q}$ interaction, and the relevant scattering process in leading order of $T/\mu_q$ for $\kappa_{bd}$ is between $bd$ and $bu$ quarks via the strong interactions.

The approximate temperature when $\kappa_{bd}$ crosses $\kappa_e$ can be calculated from Eqs.~\eqn{kappa_e_leading} and \eqn{kappa_bd_analytic}
\begin{equation}
\frac{\kappa_{e}}{\kappa_{bd}} = \frac{s^\kappa_{bu,bd}/4+s^\kappa_{bd,bd}+\tilde{s}^\kappa_{bd,bd} }{s^\kappa_{e,e}+s^\kappa_{e,bu}} \simeq 7.73 \frac{\alpha^{1/2}_s  (T/\mu_q)^3}{\alpha (T/\mu_q)^{2}} 
\label{kappa_crossover}
\end{equation}
which crosses unity at $T/\mu_q \simeq   \alpha/(7.73 \, \alpha^{1/2}_s) \simeq 10^{-3}$. 
The factor of 7.73 is a numerical constant whose only physics content is
the charges of the fermions; it is independent of $e,g,\mu_q$ and $T$.
As we anticipated, the thermal conductivity is dominated by blue down quarks at lower temperature because they do not have $\tilde{Q}$ charge and so do not feel the long-ranged (Landau-damped) $\tilde{Q}$ interaction. Their relaxation time is determined by the screened strong interactions, so the total thermal conductivity in the 2SC phase goes as $1/T$. 
This behavior is different from unpaired quark matter, in which the thermal conductivity has a constant value in the low temperature limit because of the unscreened magnetic gluon interaction (see Sec.~\ref{sec:conclusions}).

\subsection{Shear viscosity}
\label{sec:shear_viscosity}

\begin{figure}
\includegraphics[width=4.5in]{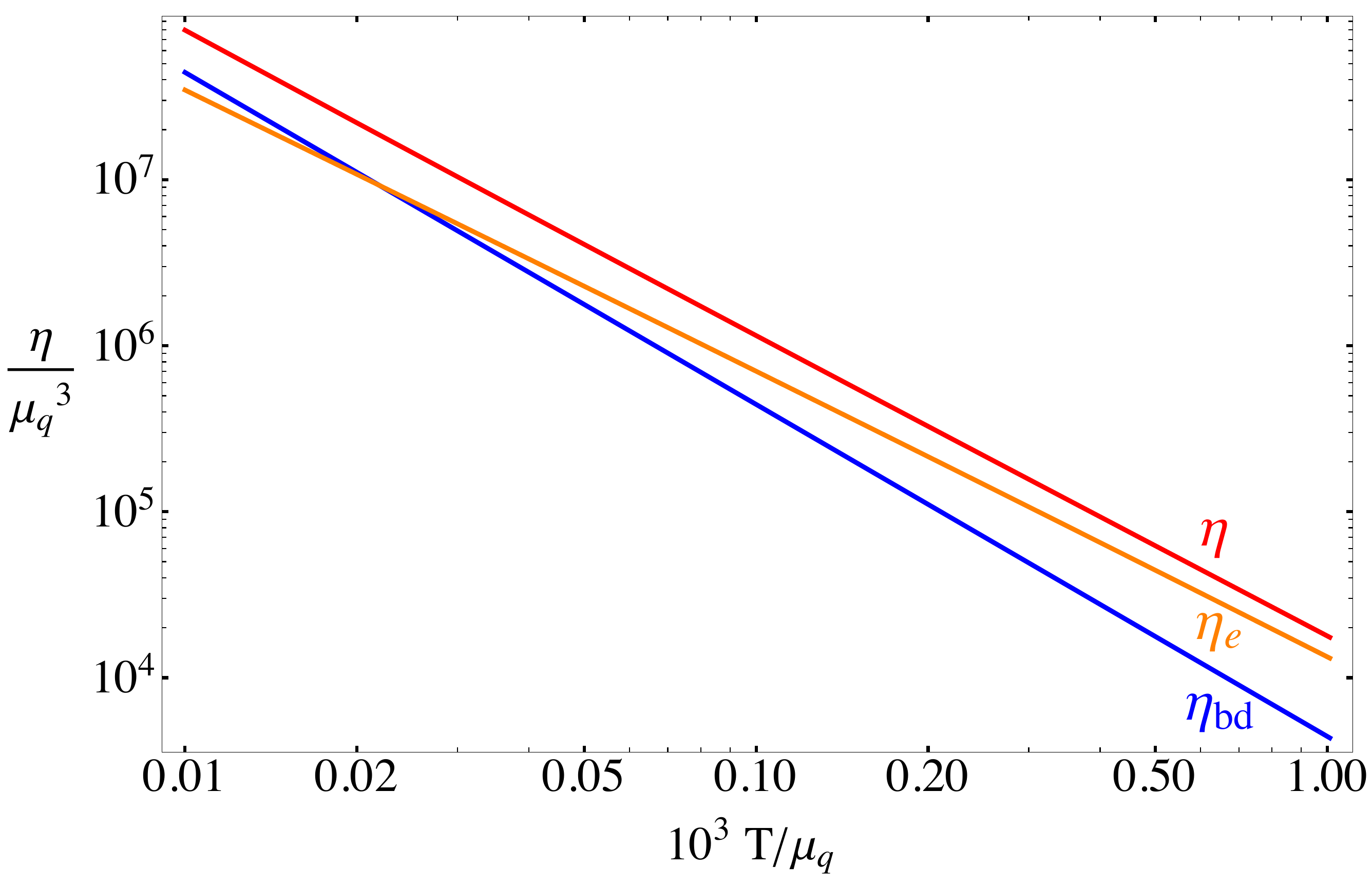}
\caption{
(Color online)
Numerical calculation of shear
viscosity as a function of temperature, taking 
$\alpha_s=1$.
In this temperature range we see the crossover from
electron domination at high temperature to blue down quark domination
at low temperature (see Sec.~\ref{sec:qualitative}).}
\label{2SC_ShearViscosity}
\end{figure}

For the shear viscosity, the transition described in Sec.~\ref{sec:qualitative} occurs only at very low temperature, so, as we now demonstrate, electrons dominate in most of the temperature range we study.
We have $\phi = p_\alpha v_\beta$ and $\gamma=5$ in
Eq.~(\ref{transport_Kinetic_2}), which then gives
\begin{equation}
\eta_i = \tau^{\eta}_i  \frac{ \mu^4_i}{5\pi^2} 
\label{shear_kinetic}
\end{equation}
where $\tau^\eta_i$ is the relaxation time for
the fermion flavor $i$. 
To calculate $\eta_i$ using Eq.~(\ref{transport_Boltzmann}) in the limit $\omega \ll q$ and $T\ll \mu_q$, we write
\begin{eqnarray}
\phi_1\cdot \left[ \tau^{\eta}_i \left( \psi_1-\psi_3 \right)+\tau^{\eta}_j \left( \psi_2-\psi_4\right) \right]
&=& 
\frac{q^2}{2}  \left( 1 -  \frac{q^2}{4 p^2_1} \right) \tau^{\eta}_i
- \frac{q^2}{2} \cos \theta  \sqrt{\left( 1 -  \frac{q^2}{4 p^2_1} \right)  \left( 1 -  \frac{q^2}{4 p^2_2} \right) } \,  \tau^{\eta}_j .
\end{eqnarray}
Using these two equations in Eq.~(\ref{transport_Boltzmann}), we have
\begin{equation}
1= \frac{9T}{4 \pi^3} \sum_j \frac{\mu^2_j}{\mu^2_i} \left( \tau^{\eta}_i  s^{\eta}_{ij} + \tau^{\eta}_j \tilde{s}^{\eta}_{ij}\right)
\label{shear_master}
\end{equation}
for each flavor $i$, where
\begin{equation}
\ba{rcl}
s^{\eta}_{ij} 
&=& \dsp
\int^\infty_{0} d\omega \left(\frac{\omega/2T}{\sinh(\omega/2T)}\right)^2
 \int^{q_M}_{0} dq  \int^{2\pi}_0 \frac{d\theta}{2\pi} 
 \left| M_{ij} \right|^2 q^2 \left(1-\frac{q^2}{4p^2_1} \right)
 \\[3ex]
 \tilde{s}^{\eta}_{ij}
 &=& \dsp
- \int^\infty_{0} d\omega \left(\frac{\omega/2T}{\sinh(\omega/2T)}\right)^2
 \int^{q_M}_{0} dq  \int^{2\pi}_0 \frac{d\theta}{2\pi} 
 \left| M_{ij} \right|^2 
q^2  \cos \theta  \sqrt{\left( 1 -  \frac{q^2}{4 p^2_1} \right)  \left( 1 -  \frac{q^2}{4 p^2_2} \right) } 
\ea
\label{s-eta}
\end{equation}
which are both symmetric in exchanging $i$ and $j$. 
We numerically evaluate both  $s^\eta_{ij}$ and $\tilde{s}^\eta_{ij}$ as functions of $T/\mu_q$ using the same parameters used in the case of the electrical conductivity. We then solve for the three relaxation times  from the three equations in Eq.~(\ref{shear_master}), and we use $ \tau^\eta_i \mu_q$ in Eq.~(\ref{shear_kinetic}) to obtain $\eta_i/\mu^3_q$. In Fig.~\ref{2SC_ShearViscosity} we show the temperature dependence of the shear viscosity, and the best fits are $\eta_{bu} /  \mu^3_q = 0.150/((T / \mu_q)^{5/3}+2490 \, (T / \mu_q)^2)$, $\eta_{bd}/\mu^3_q=0.00443/(T/\mu_q)^2$, and $\eta_e /\mu^3_q = 0.171/((T / \mu_q)^{5/3}+2.78 \,(T / \mu_q)^2)$. We note these fits can be extrapolated to arbitrary low temperature.
We can see that the electrons dominate in most of the temperature range we study, but there is a transition to the $bd$-dominated regime at low temperature, $T  \simeq 2.2 \cdot 10^{-5} \mu_q$. In the temperature range of Fig.~\ref{2SC_ShearViscosity}, the blue up contribution to the total shear viscosity is less than $0.8\%$.

We now derive approximate analytic expressions by
 assuming that the momentum transfer is much less than the
Fermi momenta ($q \ll \mu_q$). Then, as for the electrical conductivity,
we can obtain a closed form for
$s^{\eta}$ and $\tilde{s}^\eta$ using Eq.~\eqn{integral_identity_q}. In this
limit, we have $s^{\eta}_{ij} = s^{\sigma}_{ij} $ in
Eq.~(\ref{s_electrical_conductivity}) and
\begin{equation}
\tilde{s}^{\eta}_{ij} 
=
\frac{\pi^3 T}{6}
 \sum_{a= \left\{ T_8, Q\right\}}
\left[
 \frac{ Q^{a}_{i} \, Q^{a}_{j}\, Q^{X}_{i} \, Q^{X}_{j} }{\sqrt{\Pi^{aa}_l} +\sqrt{\Pi^{XX}_t}}
+ \frac{Q^{a}_{i} \, Q^{a}_{j}\, Q^{\tilde{Q}}_{i} \, Q^{\tilde{Q}}_{j}}{\sqrt{\Pi^{aa}_l} }
\right].
\end{equation}
Using $s^\eta_{ij}$ and $\tilde{s}^\eta_{ij}$ in Eq.~\eqn{shear_master}, we can solve the  three relaxation times in closed form.  These expressions are lengthy, so we only show results for $\alpha_s=1$ and $\alpha=1/137$, expanding $1/\tau^\eta_i$ to order of $T(T/\mu_q)$ to obtain 
\begin{eqnarray}
\eta_{bd} 
&\simeq& 
\frac{\mu^4_q}{5 \pi^2 }
\frac{0.111 }{T (T/\mu_q)} 
\label{eta_bd}
\\
\eta_{e} 
&\simeq& 
\frac{\mu^4_q}{5 \pi^2}
\frac{8.42}{T (T/ \mu_q)^{2/3}+ 5.56 T (T/\mu_q)}
\label{eta_e}
\end{eqnarray}
which agree with the numerical calculations to within $50\%$ for $\eta_{bd}$ and $18\%$ for $\eta_{e}$ at temperatures up to $10^{-3}\mu_q$.
As pointed out by Heiselberg and Pethick \cite{Heiselberg:1993cr}, the shear viscosity and the electrical conductivity 
vary as a different power of temperature from the thermal conductivity, because they are weighted by the momentum transfer rather than the energy transfer. 
Consequently, the ratio $\eta_{bd}/\eta_e$ becomes of order 1 at $T/\mu_q\sim 10^{-5}$ instead of $10^{-3}$ \eqn{kappa_crossover}, so most of our temperature range (shown in Fig.~\ref{2SC_ShearViscosity}) is
in the ``high temperature'' regime of Sec.~\ref{sec:qualitative}.

We write down the analytic forms of the leading terms of $\eta_e$ and $\eta_{bd}$ and identify the relevant scattering processes.  
For $\eta_e$, we can obtain the leading order of $T$ by performing an analytic calculation of the $\tilde{Q}$ interactions alone.
The solution of Eq.~\eqn{shear_master} can
then be simplified and becomes
\begin{eqnarray}
  \eta_{e} 
  & = & 
  \frac{\mu^4_{e}}{5 \pi^2} \frac{4\pi^3}{9 T (s^\eta_{e,e}+s^\eta_{e,bu})} 
=
  \frac{\mu^4_e}{5 \pi^2} \frac{8 \pi^2\Lambda^{2/3}}{3 \Gamma(8/3) \zeta(5/3) T^{5/3} \left( \bigl( Q^{\tilde{Q}}_e \bigr)^4 +\bigl( Q^{\tilde{Q}}_e \bigr)^2 \bigl( Q^{\tilde{Q}}_{bu} \bigr)^2 \right)}
  \nonumber
 \\
 &\simeq&
 \frac{\mu^4_q}{5 \pi^2} \frac{0.00231}{\alpha^{5/3} T (T/\mu_q)^{2/3}}
 \label{eta_e_leading}
\end{eqnarray}
which is Eq.~\eqn{eta_e} without the second term.
This agrees with the numerically calculated expression to $5\%$ at $T = 10^{-5}\mu_q$ and $30\%$ at $T = 10^{-3}\mu_q$.
For $\eta_{bd}$ the relaxation time is determined by
the screened $T^8$, $X$, and $Q$ interactions, but we can ignore the $Q$ interaction because $e/g$ is small. The analytic solution then becomes
\begin{eqnarray}
\eta_{bd} =
\frac{\mu^4_{bd} }{5 \pi^2} \frac{16 \pi^3}{9 T (s^\eta_{bu,bd}+4s^\eta_{bd,bd}+4\tilde{s}^\eta_{bd,bd})}
 \label{eta_bd_leading}
\end{eqnarray}
with 
\begin{equation}
s^\eta_{bu,bd}+4s^\eta_{bd,bd}+4\tilde{s}^\eta_{bd,bd} 
= 
\frac{\pi^3 T}{6} 
\left[
\frac{\bigl( Q^{T_8}_{bu} \bigr)^2\bigl( Q^{T_8}_{bd} \bigr)^2+4 \bigl( Q^{T_8}_{bd} \bigr)^4}{2\sqrt{\Pi^{T_8 T_8}_l}} + 
\frac{\bigl( Q^{X}_{bu} \bigr)^2 \bigl(Q^{X}_{bd} \bigr)^2+4\bigl(Q^{X}_{bd} \bigr)^4}{4\sqrt{\Pi^{XX}_t}} +
\frac{4 \bigl( Q^{T_8}_{bd} \bigr)^2 \bigl( Q^X_{bd} \bigr)^2 }{\sqrt{\Pi^{T_8 T_8}_l}+\sqrt{\Pi^{XX}_t}}
\right]
\nonumber
\end{equation}
which gives Eq.~\eqn{eta_bd}. 
Therefore, as in the case of the thermal conductivity, the relevant scattering process in leading order of $T/\mu_q$ for $\eta_e$ is between electrons and blue up quarks via the $\tilde{Q}$ interaction, and the relevant scattering process in leading order of $T/\mu_q$ for $\eta_{bd}$ is between $bd$ and $bu$ quarks via the strong interactions. 

The approximate temperature when $\eta_{bd}$ 
becomes equal to $\eta_e$ can be calculated from Eqs.~\eqn{eta_e_leading} and \eqn{eta_bd_leading}
\begin{equation}
\frac{\eta_{e}}{\eta_{bd}} = \frac{\mu^4_e}{\mu^4_{bd}} \frac{s^\eta_{bu,bd}/4+s^\eta_{bd,bd}+\tilde{s}^\eta_{bd,bd} }{s^\eta_{e,e}+s^\eta_{e,bu}} \simeq \frac{0.331}{2^4} \frac{\alpha^{3/2}_s  (T/ \mu_q)}{\alpha^{5/3}  (T/ \mu_q)^{2/3}}
\end{equation}
which crosses unity when $T/\mu_q  \simeq  \, (2^4 \, \alpha^{5/3})^3 /(0.331 \, \al^{3/2}_s)^3  \sim10^{-5}$. 
The factor of 0.331 is a numerical constant whose only physics content is
the charges of the fermions; it is independent of $e,g,\mu_q$ and $T$.
As noted above, this crossover temperature is much lower than that for the thermal conductivity given in 
Eq.~\eqn{kappa_crossover}. 
The reason is as follows. 
As we reduce the temperature,
the crossover occurs when the $\tilde{Q}$ interaction becomes long-ranged,
so the electron mean free path becomes short,
suppressing the electron contribution to transport
relative to that of the $\tilde Q$-neutral
$bd$ quarks (see Sec.~\ref{sec:qualitative}). However, shear viscosity and
thermal conductivity have different sensitivity to the increase in the
range of the $\tilde{Q}$ interaction.
For shear viscosity the relevant collisions are those that transfer
higher momentum
(this is related to the weight of $q^2$ in \eqn{s-eta}), so the increase in the range of the $\tilde{Q}$ interaction only has a modest impact on the mean free path, 
since the long-range interactions involve low momentum
transfer, and do not contribute much to shear viscosity. 
For thermal conductivity, the relevant collisions are those
that transfer energy (of order $T$ typically), hence the weight of $\omega^2$ in \eqn{s-kappa}, and even the low momentum
transfer interactions are able to do this.
This means that as we reduce the temperature, 
increasing the range of the $\tilde{Q}$ interaction, the contribution to
shear viscosity from electrons is only moderately suppressed relative
to that from $bd$ quarks, whereas the  contribution to
thermal conductivity from electrons is heavily suppressed relative to
that from $bd$ quarks. Consequently, for shear viscosity we have to
go to much lower temperatures in order to reduce the electron contribution
to the same level as the $bd$ contribution.

\hide{
The crossover temperature is determined by the scattering integrals $s_{ij}$ and $\tilde s_{ij}$ (see \eqn{s-kappa}, \eqn{s-eta}). In the limit $q \ll \mu_q$, $s_{ij}$ takes the form
\begin{equation}
s_{ij}=
\int^\infty_0 d \om \left( \frac{\omega / 2T}{ \sinh (\omega/2T)} \right)^2 \int^\infty_0 dq f(\om,q) |M_{ij} |^{2}
\end{equation}
where for thermal conductivity the weight is $f=\omega^2$ and for the shear viscosity the weight is $f=q^2$. 
For $\tilde{s}_{ij}$, we have a similar expression but with some $\theta$ dependance. 
Typical momentum transfer is of order $q \sim \sqrt{ \Pi^{aa} }$, and it depends on the interactions: that for the strong interaction, $q_{typ,g}$, is of order $\sim g \mu_q$ while that for the $\tilde{Q}$ interaction, $q_{typ,e} $, is of order $\sim \Lambda^{2/3} T^{1/3} \sim e^{2/3} \mu^{2/3}_q T^{1/3}$, which is temperature dependent due to the Landau damping. From the analytical expressions, we have seen that the electron contributions and the blue down contributions in the leading order of $T$ are inversely proportional to the scattering integrals with the $\tilde{Q}$ interaction and the strong interactions, respectively.  Noting that $|M_{ij}|^2 \sim (Q^a_i)^2 (Q^a_j)^2 /|(q^2 + \Pi^{aa})|^2$, the electron and blue down contributions to the thermal conductivity are (see \eqn{kappa_e_leading}, \eqn{kappa_bd_analytic})
\begin{equation}
\kappa_{e} \sim T^2 \frac{q^3_{typ,e}}{e^4 \omega^3_{typ}}  \,\,\,\,\,\,\,\,\,\,\,\,\,\,\,\,\,\,\,\, \kappa_{bd} \sim T^2 \frac{q^3_{typ,g}}{g^4 \omega^3_{typ}}
\end{equation}
where typical energy transfer is always of order $\omega_{typ} \sim T$ because of the Boltzmann factor.  Similarly for the shear viscosity, we have (see \eqn{eta_e_leading}, \eqn{eta_bd_leading})
\begin{equation}
\eta_{e} \sim \frac{\mu^4_e}{T} \frac{q_{typ,e}}{e^4 \omega_{typ}}  \,\,\,\,\,\,\,\,\,\,\,\,\,\,\,\,\,\,\,\, 
\eta_{bd} \sim \frac{\mu^4_{bd}}{T} \frac{q_{typ,g}}{g^4 \omega_{typ}}.
\end{equation}
The dependance of typical momentum and energy transfers differs between the thermal conductivity and the shear viscosity because of the weight $f$ in the scattering integrals. In both cases, the electron contribution has a higher power of temperature than the blue down contribution because of the temperature dependance of $q_{typ,e}$ due to the Landau damping effect, and this effect is larger for the thermal conductivity because the power of typical momentum transfer is higher. As a result, the electron contribution decreases faster than the blue down contribution as we lower the temperature for the thermal conductivity than for the shear viscosity, and therefore, the crossover temperature for the shear viscosity is lower than that for the thermal conductivity.
Note, however, that because there are more blue down quarks than electrons, the crossover temperature for the shear viscosity is higher by a factor of $(\mu^4_{bd}/\mu^4_e)^3 \simeq 4 \cdot 10^3$ than it would be in the hypothetical case when they have the same chemical potential.   
}

\section{Vortex-fermion scattering contribution to the transport}
\label{sec:fluxtubes}

If the 2SC core of a neutron star forms in the presence of a
magnetic field, it will be penetrated by the $\tilde Q$ component of
the field, but behave as a type-II superconductor with respect to
the $\X$ component, so the $X$ flux is concentrated into 
``color-magnetic'' flux tubes \cite{Alford:2010qf}. 
It is not yet clear whether these
flux tubes are energetically stable, but in this section we estimate their
possible contribution to transport via
scattering of ungapped fermions off the flux tubes. 
Because the density of flux tubes is independent of temperature,
but the density of ungapped fermions decreases with $T$,
flux-tube scattering will eventually dominate transport at sufficiently low
temperatures. 

For a given transport coefficient $\xi =\{\sigma, \kappa, \eta \}$ the relaxation time $\tau^\xi_i$ of a fermion of type $i$ is inversely proportional to the sum of the inverse relaxation times associated with the different scattering channels:
\begin{equation}
\frac{1}{\tau^\xi_i} =  \frac{1}{\tau^\xi_{i,v}} + \sum_{j} \frac{1}{\tau^\xi_{ij}} 
\end{equation}
where $1/\tau^\xi_{i,v}$ and $1/\tau^\xi_{ij}$ are the fermion-vortex and fermion-fermion relaxation rates, respectively.
In order to give a simple estimate, we assume that the fermion-vortex and fermion-fermion relaxation rates are decoupled, so $\sum_j 1/\tau^\xi_{ij}$ is simply the inverse of the relaxation times which we have computed in the previous section. It is then clear from this expression that the fermion-vortex scattering process only increases the total relaxation rate and thus only suppresses the transport coefficient. 

The vortex-fermion contribution has been discussed in \cite{Alford:2010qf}, and here we give a brief explanation of the result. 
The $X$-flux tubes have area density $n_v = B/\Phi_X$, where $\Phi_X = 6 \pi/e$. 
The ungapped fermions will scatter off the color-magnetic flux tubes via the Aharonov-Bohm effect.
The cross section of the Aharonov-Bohm scattering is proportional to $ \sin^2(\pi\tilde\beta_i) /\mu_i$, where $\tilde\beta_i$ is a measure of the Aharonov-Bohm interaction of the fermion of type $i$ with the $X$-flux tube. The $bu$ quark and the electron have the same factor $\sin^2(\pi\tilde\beta_i) \simeq \pi^2\alpha^2/\alpha_s^2$, while for the $bd$ quark this factor is zero \cite{Alford:2010qf}. Because the vortex does not interact with the blue down quark, it does not affect the blue-down quark contributions to the transport coefficients.  

\noindent\underline{Thermal conductivity}. 
From the previous section, we have found that the dominant contribution to the thermal conductivity is from blue down quarks, which do not interact with vortices.  Therefore, the vortex scattering process only suppresses the subdominant contributions and does not affect the total thermal conductivity in the temperature range we have considered in the previous section.

\noindent\underline{Electrical conductivity and shear viscosity}.
In the absence of vortices, we have found in the previous section that the dominant contribution to both the electrical conductivity and the shear viscosity is from electrons, and the most relevant interaction for the relaxation rate is the $\tilde{Q}$ interaction. From Eqs.~\eqn{sigma_kinetic} and \eqn{electrical_conductivity} for the electrical conductivity and from Eqs.~\eqn{shear_kinetic} and \eqn{eta_e_leading} for the shear viscosity, we can read off the relaxation rates of the electron for the two transport coefficients and write them as
\begin{equation}
\sum_j \frac{1}{\tau^{\xi}_{e,j}}
=
c_\xi \, \alpha^{5/3} \frac{T^{5/3}}{\mu^{2/3}_q}
\label{relaxation_rate_e_j}
\end{equation}
where $\xi = \{ \sigma, \eta \}$ and for each $\xi$, $c_\xi$ is a numerical constant of order $10$, which depends on the charges of the fermions.  
According to \cite{Alford:2010qf}, the momentum relaxation rate for the electron-vortex scattering is 
\begin{equation}
\frac{1}{\tau_{e,v}}  = \frac{\pi^{3/2} \alpha^{5/2}}{3 \alpha^2_s} \frac{B}{\mu_e}.
\label{relaxation_rate_e_v}
\end{equation}
We expect, as is the case for the fermion-fermion relaxation rates \cite{Heiselberg:1993cr}, that the electron-vortex relaxation rates for electrical conductivity and shear viscosity are the same as the momentum relaxation rate up to a constant of order $1$.  
Electron-vortex scattering becomes important when its rate \eqn{relaxation_rate_e_v} becomes comparable to the fermion-fermion relaxation rate \eqn{relaxation_rate_e_j}.
Taking $\alpha_s=1$ and assuming typical chemical potential  $\mu_q = 400 \, \mbox{ MeV} $ and the lowest possible temperature in neutron stars to be $T=10^7 \, \mbox{K}$, we find that the ratio of the rates becomes unity when the magnetic field reaches
\begin{equation}
B \sim  10^{12} \, \mbox{G} \, \left(\frac{T}{10^7 \, \mbox{K}} \right)^{5/3} \left(\frac{\mu_q}{400 \, \mbox{MeV}} \right)^{1/3}.
\end{equation}

From the above estimates, we conclude that the presence of the vortices in realistic values of the external magnetic field can lower the transport coefficients we have computed in the previous section. Therefore, performing more complete computations of transport coefficients with the presence of vortices may be necessary if the vortex in the 2SC phase turns out to be stable.

\section{Conclusions}
\label{sec:conclusions}

We have calculated the electrical conductivity, thermal conductivity, and shear viscosity of quark matter in the 2SC phase using the linearized Boltzmann equation in the relaxation time approximation. We have relied on perturbation theory and used the leading order in $\alpha$ and $\alpha_s$ for the scattering matrix element. 
For the numerical computations, we have  assumed that the energy transfer $\omega$ is much smaller than the momentum transfer $q$ (the static limit) and have taken the leading order in $\omega/q$. 
In the temperature range $10^{-5} < T /\mu_q < 10^{-3}$, this approximation is good because the characteristic energy transfer is temperature, while the characteristic momentum transfer is the Debye screening mass.  
The results are shown in Figs.~\ref{2SC_ElectricalConductivity}, \ref{2SC_ThermalConductivity}, \ref{2SC_ShearViscosity} for $\alpha_s =1$.
% We have found that all transport coefficients in the 2SC phase are
% quantitatively different from those in the unpaired quark matter phase, 
% and in some case, power laws of temperature are also different.
For physical insight, we have obtained approximate analytic results by further assuming that the momentum transfer is much smaller than the quark chemical potential, $q \ll \mu_q$.   For the electrical \eqn{electrical_conductivity}
and thermal \eqn{kappa_bd_analytic} conductivities, the analytic results of the leading fermion contributions agree with the numerical results of the leading fermion contributions to within 22\% and 6\%, respectively.
For the shear viscosity, the leading (electron) contribution 
\eqn{eta_e} agrees with the numerical result of the leading fermion contribution to within about 18\% over the relevant temperature range.

The general picture of transport in the 2SC phase is that it occurs
via the ungapped fermions, which are the blue up quark, the blue down quark, and the electron. The electron contribution dominates at higher temperature because electrons do not feel the strong interaction, only the electromagnetic interaction, and so have longer relaxation times than the ungapped quarks. However, at low temperature the $\tilde{Q}$ interaction becomes long-ranged because it is Landau damped (not Meissner-screened), and this compensates for its small inherent coupling. The $\tilde{Q}$-neutral blue down quark therefore dominates transport at low temperatures, because its interactions, although strong, are screened.

\noindent\underline{Thermal conductivity}:
the crossover from blue-down to electron
domination occurs at $T/\mu_q\sim \alpha/7.7\sim 10^{-3}$, so most of
the temperature range of interest for neutron stars is in the 
blue-down-dominated regime where $\ka\sim 1/T$.

\noindent\underline{Shear viscosity}: the crossover from blue-down to electron
domination occurs at $T/\mu_q\sim  10^{-5}$, so electrons are dominant down
to $T\sim 10$\,keV.
The crossover temperature for the shear viscosity is much smaller than for the thermal conductivity because the relevant collisions for shear viscosity are those that transfer higher momentum, so the increase in the range of $\tilde{Q}$ interaction has a smaller impact on the mean free path since the long-range interactions involve low momentum transfer. See the end of Sec.~\ref{sec:shear_viscosity}.

\noindent\underline{Electrical conductivity}: this
is a special case because the transported quantity is 
$\tilde{Q}$ charge, so the blue down quarks, which
are $\tilde Q$ neutral, do  not 
contribute to the electrical conductivity. The electron contribution 
therefore dominates over the entire temperature range.

Other possible excitations that might contribute to the transport coefficients include the color-magnetic flux tubes and gluons in the unbroken gauge sector.  Flux tubes are discussed in Sec.~\ref{sec:fluxtubes}, where we have argued that at sufficiently low temperature and high magnetic field, the vortex-fermion scattering via the Aharonov-Bohm effect may suppress the electron contributions to the electrical conductivity and shear viscosity.

We now argue that $SU(2)_{rg}$ gluons do not contribute to the transport coefficients in
2SC quark matter. The glue sector of the unbroken $SU(2)_{rg}$ gauge
theory has a confinement scale $\Lambda'_{QCD}$ which may be in the
keV range, or as high as about 10\,MeV and a coupling 
 $\alpha'_s \simeq (\pi \alpha_s/2)^{1/2} \Delta/\mu_q$ which is smaller than $\alpha_s$ because of the partial screening
of the Cooper pairs \cite{Rischke:2000cn}.
If $T\ll \Lambda'_{QCD}$ then the gluons are confined into glueballs with
mass of order $\Lambda'_{QCD}$, so their contributions to transport are
exponentially suppressed.
If  $T\gtrsim \Lambda'_{QCD}$ then the theory is deconfined, and the
gluons can contribute to the transport coefficients.
From dimensional analysis, we can estimate that the $SU(2)$ gluon 
contributions to
the thermal conductivity and the shear viscosity are $\kappa_{\rm{glue}} \sim
T^2/\alpha'^2_s$ and $\eta_{\rm{glue}} \sim T^3/\alpha'^2_s$ where $1/\alpha'^2_s$
comes from the scattering amplitude of the gluons.  Comparing the gluon
contributions with the blue down quark contributions for $\alpha_s =1$ (see
Eqs.~\eqn{kappa_bd_analytic} and \eqn{eta_bd_leading}), we have
$\kappa_{\rm{glue}}/\kappa_{bd} \sim (T/\mu_q)^3 (\mu_q /\Delta)^2 $ and
$\eta_{\rm{glue}}/\eta_{bd} \sim (T/\mu_q)^5 (\mu_q /\Delta)^2$, which are both much
less than 1 because $T / \mu_q \lesssim10^{-3}$ and $\mu_q /\Delta < 2 \mu_q/ \mu_e \lesssim 3.5$ for 2SC quark matter (see end of Sec.~\ref{sec:relevant_excitations}). The electrical conductivity has a
contribution from the gluons because, like the blue up quark, they carry
a $\tilde{Q}$ charge arising from their color $T_8$ charge. However,
dimensional analysis shows that their contribution is $\sigma_{\rm{glue}} \sim
e^2T/\alpha'^2_s$, which is also negligible compared to the electron
contribution \eqn{electrical_conductivity}.

It is interesting to compare our results with the transport properties of two
flavor {\em unpaired} quark matter.  
Transport in unpaired quark matter is dominated by electrons and their
electromagnetic scattering off gapless quarks. This is because there is no
Meissner screening of the gluons; they are Landau damped like the photon, so both gluon and photon interactions become long-ranged at low temperature, and there is no crossover to a regime where short-ranged strong interactions dominate transport. 
We therefore expect that the transport coefficients of unpaired 2-flavor quark matter are similar to those we calculated as the electron contribution to 2SC quark matter [\eqn{electrical_conductivity}, \eqn{kappa_e_leading}, and \eqn{eta_e_leading}]. 
The electron contributions to the transport coefficients of unpaired quark matter can be easily computed. After performing calculations similar to those in Sec.~\ref{sec:ff_scattering}, we can show that the transverse component of the photon self-energy and the electron chemical potential in unpaired quark matter are  $2.34  i \omega \Lambda^2/q$ and $0.219 \mu_q$, respectively.  Note that the chemical potential of the electron in unpaired quark matter is smaller than in 2SC quark matter. Using these values in  \eqn{electrical_conductivity}, \eqn{kappa_e_leading}, and \eqn{eta_e_leading}, we find the electron contributions to the transport coefficient in unpaired quark matter are $0.0070 \sigma_e$, $1.7 \kappa_e$, and $0.022 \eta_e$. 
The electrical conductivity and shear viscosity are much smaller because of their $\mu_e$ dependance.
We conclude that the electrical conductivity and shear viscosity, which are dominated by electrons,
have similar expressions in 2SC quark matter and unpaired 2-flavor quark matter in the temperature range that we studied.
However, the thermal conductivity of 2SC matter is dominated by $bd$ quarks, and rises as $\mu_q^3/T$ at low temperature, whereas in unpaired 2-flavor quark matter it is dominated by electrons and tends to a constant value of order $\mu_q^2$.  It is an interesting future project to compute the transport properties of unpaired quark matter numerically and give more rigorous comparison with 2SC quark matter.

One natural generalization of our results would be to analyze the 2+1 flavor case, where strange quarks help to ensure neutrality, and $\mu_e$ is much
smaller and depends on the strange quark mass. 
This will affect the dominance of electrons.
Another application would be a more careful treatment of 
2-flavor and 2+1-flavor unpaired quark matter (to our knowledge,
only 3-flavor unpaired quark matter has been treated in the literature
\cite{Heiselberg:1993cr}
and it is a special case because of the absence of electrons).
One could then go on to study applications of these results to the
observables on neutron stars. 
Shear viscosity plays an important role in the spindown behavior of
neutron stars, since it is one of the dissipation mechanisms that damps
``r-modes". Without sufficient damping, r-modes would arise spontaneously
in fast-spinning neutron stars, spinning them down via emission of
gravitational radiation \cite{Alford:2010fd,Andersson:1998}. 
Thermal conductivity is the key microscopic 
quantity that controls  macroscopic thermal transport and equilibration in the dense cores of young (less than a few hundred years old) isolated  neutron stars and in accreting transient X-ray sources. 
Hybrid compact stars with 2SC matter may relax thermally 
on timescales that are different from those of their hadronic counterparts
and this can be tested observationally. 
Finally, electrical conductivity of 2SC matter determines the timescale for the 
decay of the component of magnetic field which is not frozen in the color-magnetic 
flux-tubes. Addressing this problem requires (in addition to conductivities of 
various phases) the knowledge of large-scale structure of the magnetic field and, 
therefore, the current distribution within the star. The putative decay of the magnetic 
fields can be tested, for example, with the models of secular evolution of pulsars in 
the $p-\dot p$ diagram.

\section*{Acknowledgements}

We thank Toru Kojo, Dirk Rischke, Andreas Schmitt, and Kai Schwenzer for
useful discussions. This research was supported in part by the Offices
of Nuclear Physics and High Energy Physics of the U.S.~Department of
Energy under contracts \#DE-FG02-91ER40628, % Wash U theory
\#DE-FG02-05ER41375, % Mark DoE
the Sofja Kovalevskaja program of the Alexander von Humboldt Foundation, the Bielefeld Young Researcher's Fund, and the Deutsche 
Forschungsgemeinschaft (Grant No. SE 1836/3-1).

% \appendix  % Use \appendix* if there is only 1 appendix

% start comment

% \appendix*

% end comment

% Override the revtex href command in order that the JHEP bib style
% will work properly:
\renewcommand{\href}[2]{#2}
% macros used by ADS Database BiBTeX entries:
\newcommand{\apjl}{Astrophys. J. Lett.}
\newcommand{\mnras}{Mon. Not. R. Astron. Soc.}

\bibliographystyle{JHEP_MGA}
\bibliography{2SC_transport} 
\end{document}